\documentclass[12pt,preprint]{aastex}

\def\plotfiddle#1#2#3#4#5#6#7{\centering \leavevmode
\vbox to#2{\rule{0pt}{#2}}
\includegraphics{#1}}

\shorttitle{Investigating Variable Protostars using 2-4$\mu$m Spectra}
\shortauthors{Beck}

\begin{document}


\title{Investigating the Nature of Variable Class I and Flat Spectrum Protostars Using 2-4$\mu$m Spectroscopy}


\author{Tracy L. Beck\altaffilmark{1,2}}

\affil{1 - Gemini Observatory, Northern Operations, 670 N. A'ohoku Pl. Hilo, HI 96720}
\affil{2 - Visiting Astronomer at the Infrared Telescope Facility, which is operated by the University of Hawaii under Cooperative Agreement no. NCC 5-538 with the National Aeronautics and Space Administration, Science Mission Directorate, Planetary Astronomy Program.}




\begin{abstract}

In this study I present new K and L$'$-band infrared photometry and 2-4$\mu$m spectra of ten Class I and flat spectrum stars forming within the Taurus dark cloud complex.  Nine sources have H$_2$ {\it v}=0-1 S(1) emission, and some show multiple H$_2$ emission features in their K-band spectra.  Photospheric absorptions characteristic to low mass stars are detected in five of the targets, and these stars were fit with models to determine spectral type, infrared accretion excess veiling (r$_K$ and r$_{L'}$) and dust temperatures, estimates of visual extinction and characteristics of the 3$\mu$m water-ice absorption.  On average, the models found high extinction values, infrared accretion excess emission with blackbody temperatures in the 900-1050K range, and 3$\mu$m absorption profiles best fit by water frozen onto cold grains rather than thermally processed ice.  Five techniques were used to estimate the extinction toward the stellar photospheres; most gave vastly different results.  Analysis of emission line ratios suggests that the effect of infrared scattered light toward some protostars should not be neglected.  For stars that exhibited Br$\gamma$ in emission, accretion luminosities were estimated using relations between L$_{acc}$ and Br$\gamma$ luminosity.  The young stars in this sample were preferentially chosen as variables, but they do not have the accretion dominated luminosities necessary to put them in their main stage of mass-building.  The characteristics of the 2-4$\mu$m spectra are placed in the context of existing multi-wavelength data, and five of the stars are catagorized as reddened Class IIs or stars in transition between Class I and II, rather than protostars embedded within massive remnant envelopes.

\end{abstract}




\section{Introduction}

Ever since the identification of T Tauri stars as young solar analogs (Ambartsumian 1947; 1949), astronomers have sought to better understand the youngest phases of stellar evolution.  The youngest optically revealed stellar objects (YSOs) are still obscured by their natal envelopes and ambient cloud material, and hence are very faint in visible light.  Understanding the stellar and circumstellar properties of these young stars has proven feasible with the current generation of sensitive instrumentation and large aperture telescopes.  The Class I protostars represent the youngest YSOs which are presently identified at optical and near-infrared wavelengths.

 In the accepted classification scheme for early stellar evolution, Class I, or protostellar T Tauris, are embedded young stars that are intermediate between extremely young stellar embryos (Class 0) and the more familiar Classical T Tauri Stars (CTTSs, Class IIs; Adams, Lada \& Shu 1987; Lada 1987).  The definition of these classes came from the shape of the spectral energy distribution (SED) from 2-25$\mu$m.  The spectral index, $\alpha$ = dlog($\lambda$F$\lambda$)/dlog($\lambda$), represents the shape of the SED. Sources that have $\alpha>$0.0 are designated Class I with rising SEDs in the infrared, and those with $\alpha\approx$0.0 are flat spectrum objects (Adams, Lada \& Shu 1987; 1988).  The SED shapes of Class Is are well modeled as star+rotating disk systems with massive infalling envelopes and outflows (Kenyon et al. 1993a; 1993b, Kenyon \& Hartmann 1995; Whitney et al. 1997; 2003; Terebey et al. 2006).  It has become apparent from both the models and observations that a degeneracy in the spectral index exists between embedded protostars and more evolved but obscured T Tauris, e.g., CTTSs with disks viewed edge-on (Whitney et al. 1997; 2003; Chiang \& Goldreich 1997; 1999; Hogerheijde \& Sandell 2000).  Investigations in optical through millimeter regimes have sought to better define the differences between Class I stars and reddened Class IIs (Motte \& Andr\'e 2001; Park \& Kenyon 2002; White \& Hillenbrand 2004; Andrews \& Williams 2005; Eisner et al. 2005). 

One of the outstanding discrepancies between the theory and observation of Class I protostars is the ``luminosity problem'' first described by Kenyon et al. (1990).  If mass in an infalling protostellar envelope is accreting efficiently onto the stellar surface instead of only onto the circumstellar disk, then the luminosity from mass accretion should dominate the stellar photosphere by a factor of ten or more.  However, for the Class I stars that have been investigated, the mass accretion luminosities do not differ significantly from those of Class IIs (Gullbring et al. 1998; Muzerolle et al. 1998; White \& Hillenbrand 2004).  This has led to speculation that either Class I protostars are not in their main accretion phase and their mass was accumulated at a younger epoch (White \& Hillenbrand 2004), or that eruptive events of enhanced mass accretion in FU Ori-like outbursts are responsible for building the bulk of the stellar mass (Kenyon et al. 1990; Calvet et al. 2000).  While the flux of some Class Is is known to vary (Leinert et al. 2001; Park \& Kenyon 2002; Forbrich, Preibisch \& Menten 2006), little is known about the cause and character of variability in most embedded protostars.

Spectral characteristics in the near-infrared provide important opportunities to study the environments and natures of YSOs (Greene \& Lada 1996).  The K-band (2.0-2.4$\mu$m) spectra of Class Is typically have continuum emission that is flat or rising steeply toward longer wavelengths, and are often featureless aside from emission in atomic or molecular hydrogen transitions which indicate ongoing mass accretion and outflow activity, respectively (Greene \& Lada 1996; 1997; Prato \& Simon 1997; Muzerolle et al. 1998; Davis et al. 1994; Gomez et al. 1997).  On average, Class Is have greater K-band infrared accretion excesses than the more evolved CTTSs, and stellar photospheric absorption features characteristic to their spectral types are often veiled to undetectability by the excess emission (Greene \& Lada 1996; 1997; Luhman \& Rieke 1999; Doppmann et al. 2005; Ishii et al. 2004). 

2-4$\mu$m spectroscopy provides many useful indicators to determine spectral type, accretion and outflow activity, and the characteristics of material along the line of sight to young stars (Whittet et al. 1988; Greene \& Lada 1996; 1997; Prato \& Simon 1997; Teixeira \& Emerson 1999; Muzerolle et al. 1998; Muzerolle et al. 2003).  Yet, this wavelength region has not been used extensively for Class Is.  In this work, I present 2-4$\mu$m spectroscopy of a sample of ten protostars in the Taurus dark cloud complex.  These stars were chosen because they had Class I or flat SED shapes (Kenyon \& Hartmann 1995) and had evidence for near infrared flux variability (Park \& Kenyon 2002).  In \S2 the observations are discussed.  In \S3 I present the spectra and the models used to characterize the stars and \S4 and \S5 discuss extinction and accretion properties, respectively.  In \S6 I place the results in the context of existing data and briefly discuss each star.

\section{Observations and Data Analysis}

The data for this project were obtained at the NASA Infrared Telescope Facility (IRTF) 3 meter telescope at Mauna Kea on UT 2003 Dec. 5 and 6.  K and L'-band images and 2-4$\mu$m cross-dispersed spectra were acquired.  During the observations, the sky was photometric and the seeing was 0.$''$6-0.$''$8.  The facility near infrared imager and spectrometer ``SpeX'' was used to take the data (Rayner et al. 2003).   SpeX consists of a 1024$\times$1024 InSb array, Bigdog, which has a plate scale of 0.$''$15/pixel and is used for spectroscopic observations, and a 512$\times$512 InSb array, Guidedog, which has a 0.$''$12/pixel plate scale and serves as a guiding/imaging camera.   The targets for this project were chosen because they are potential variable stars; they showed $>$0.15 mag of deviation from their mean K-band magnitudes in the study of Park \& Kenyon (2002).  Most of the young stars were not spatially extended by more than FWHM$\sim$2.$''$2 (2 pixels) in the study of Park \& Kenyon (2002).  

Table 1 presents the list of the stars observed for this project.   Also included in Table 1 are the RA and Dec. for the targets, the SED classifications, the standard deviation of multiple measurements of the magnitude, $\sigma_K$, which provides an estimate of the stellar variability level, and the K and L$'$ magnitudes determined from the data presented in this work. (Kenyon \& Hartmann 1995; Park \& Kenyon 2002; White \& Hillenbrand 2004).  For this study, stars with 2-25$\mu$m spectral indices of less than 0.3 are identified as Flat spectrum sources rather than true Class Is (Table 1; Adams, Lada \& Shu 1988).  Photometry was obtained for IRAS 04108+2803A because it lies in the same image as 04108+2803B, but no spectroscopy was acquired for this source because its value for $\sigma_K$ did not classify it as a potential variable star (Park \& Kenyon 2002).

\subsection{Imaging}

For the 11 stars in Table 1, images were obtained using the Guidedog camera with the K and L$'$ filters.  The integration times were 160 seconds at K and 80 seconds at L$'$.  This total time was derived from summing 40 images, each individual image was an 0.8 second integration with 5 coadds in the K-band, and 0.2 second integration with 10 coadds at L$'$.  For sky subtraction, the data were obtained by cycling through an a/b beam-switch pattern using telescope offsets of 15$''$.  Photometry for each of the sources was done using a 5$''$ circular aperture.  The photometric standard 16 Tau was observed 3 times per night for K-band flux calibration and 2 times for the L$'$-band; multiple photometric standards were not observed.  Estimated uncertainties are $\sim$0.08 and $\sim$0.2 magnitudes in the K and L$'$-bands, respectively.  Columns 6 and 7 of Table 1 present the final K and L$'$ magnitudes of the stars. 

\subsection{Spectroscopy}

The spectra were obtained using the Bigdog array with a 0.$''$5 slit in the long wavelength cross-dispersed setting, LXD1.9, which provides simultaneous spectral coverage of the 1.9-4.1$\mu$m region.  In order to correct atmospheric features at 3.0-3.4$\mu$m, the spectra were obtained by observing multiple 30 second exposures with a beam-switch of 7.$''$5 that kept the star in the slit at all times.  For the success of telluric correction, observations of an A0 spectral type star were made within $<$20 minutes in time and less than 0.2 difference in airmass of each science target.

The data were reduced using the ``SpexTool'' data reduction package provided by the IRTF (Cushing et al. 2004). The spectra were extracted using an optimal extraction with a PSF fitting radius of 1'' and an extraction aperture of 1.''2. The SpexTool tasks xtellcor, xmergeorders and xcleanspec were used for telluric correction, to merge the orders into a single spectrum, and to clean bad pixels.  The photosperic hydrogen features present in the A0 spectral telluric calibrator were removed using the ``xtellcor'' data reduction routine that uses the deconvolution method described by Vacca et al. (2003).  In all cases, good correction of the hydrogen photospheric features in the A0 stars was obtained using this method.    To test the success of the telluric correction, multiple A0 stars were run through ``xtellcor'' as the science targets.  Correction of the spectra for these stars was done using A0 stars observed nearby in time and airmass.  The resulting telluric correction was good in the 2.40-2.44$\mu$m region where H$_2$ Q-branch emission in the spectra of young stars can be adversely affected; remnant telluric features were typically not seen above the noise in the continuum in the A0 stars.  The fluxes derived from the K and L$'$ images of the young stars were used to calibrate the spectra.  

Observations of thirteen late type dwarf and giant stars with spectral types of K1 through M6 were acquired for comparison with the spectra of the young stars.  These spectral type templates were necessary for use in spectral model fits.  The stars were chosen based on their spectral types listed in the Hipparchos catalog, and their brightness in the infrared (all had K$<$7.5). These spectra were acquired either on UT 2003 Dec. 5 and 6 at the same time as the observations of the young stars, or in a previous session of observing for a separate project (UT 2002 May 30 - June 1).  Observations of the spectral templates were acquired with the same instrument configuration as the young stars, with three to five a/b beam-switched cycles (depending on the brightness of the star) of 30 second individual exposures.  The spectra were extracted and corrected for telluric features following the method described above for the young stars.  Additional spectral type comparison stars were also obtained from the SpeX library of stellar standards available online through the IRTF webpage (Cushing et al. 2003){\footnote{Spectral template standards obtained with the SpeX instrument are available at: http://irtfweb.hawaii.edu/spex}}. 

The final spectra of the Class I and flat spectrum stars have resolution R $\sim$ 1100 in the K-band.  Figure 1a through 1j present the 1.95-4.07$\mu$m spectra of the 10 protostars and the lower panels show a zoomed view of the K-band region. 




\section{Results}

For most of the targets, the measured K-band magnitudes are comparable to those found from past studies and values reported by 2MASS.  A few stars are 0.1-0.2 magnitudes brighter than 2MASS, but this is perhaps expected from the profile of the shorter wavelength K$_s$ filter used to acquire the 2MASS data and the reddened spectral shape of most of the targets.  One source, IRAS 04239+2436, has brightened to 9.88 mag compared to the average of 10.33 from Park \& Kenyon (2002).  This was one of the more variable stars in the sample (Table 1), but given the inherent uncertainties the photometry presented here and that of Park \& Kenyon (2002), the variation in magnitude for IRAS 04239+2436 is only at a $\sim$3$\sigma$ level of significance.  It is also difficult to reach any new conclusions on the variability of these stars because of differences in filter profiles, photometric apertures, and plate scales used for our respective works.  None of the stars were spatially resolved with respect to the images of the flux standard star.  The spatial pixels were too small and the total integration times too short to examine with confidence the character of low surface brightness nebulosity in the vicinity of these young stars.





In Figure 1a through 1j, the spectra of the stars are plotted; the upper panel presents the 2-4$\mu$m region and the lower shows a close-up of the K-band.    Table 2 provides a summary of the emission and absorption features in the 2-4$\mu$m region that have been detected in these spectra and are used to better understand young stars.  Also included in Table 2 is a note on the information the line species provides.  Young, unveiled, low mass stars have absorption features in their K-band spectra characteristic to their spectral type (Wallace \& Hinkle 1997; Greene \& Lada 1996; e.g. Figure 1b).  Photospheric absorption features are detected in five of these stars.  

In low mass protostars, molecular hydrogen emission typically arises from regions where outflows plow into the ambient ISM and shock the local material.  Emission from H$_2$ {\it v}=1-0 S(1) at 2.12$\mu$m was detected in all but one of the stars (IRAS 04181+2654B), and a few stars showed emission from multiple molecular hydrogen transitions in the K-band (Figure 1g).  Unfortunately, the signal-to-noise (S/N) of the L-band spectra were too low to detect H$_2$ features present in this range.  Additionally, the H$_2$ {\it v}=2-1 S(1) emission at 2.24$\mu$m was not seen in any of the spectra.  Atomic hydrogen emission arises from the magnetospheric accretion of material onto the central star, features in the infrared are surrogates to the more familiar H$\alpha$ emission in the optical (e.g. Figure 1e; Prato et al. 1997; Muzerolle et al. 1998).  In addition to Br$\gamma$ in the K-band spectra, strong emission from Br$\alpha$ (4.05$\mu$m) is also detected in many of the stars.  Table 3 presents the line fluxes of the emission features detected in the spectra and 3$\sigma$ detection limits are included where the emission features were not detected.  

 Beyond a certain threshold extinction, ices frozen onto interstellar or circumstellar dust grains give rise to the solid-state absorptions from O-H and C-H stretch-mode oscillations at 3.05 and 3.47$\mu$m, respectively (e.g. Figure 1i; Whittet et al. 1988, Teixeira \& Emerson 1999, Brooke et al. 1999).  The 3$\mu$m water ice absorption in the L-band spectra are detected at varying levels of confidence in all but one of the young stars (IRAS 04264+2433).  This absorption can arise from ices frozen onto grains in the circumstellar disk, natal envelope, or from ambient material in the Taurus dark clouds.  The shape of the water ice absorption profile is temperature dependent, a narrow, more sharply peaked shape is indicative of warm, annealed ice and a broad, rounded profile traces cooler dust (Gerakines et al. 1995; 1996).  Investigating the profile shape of the water feature can provide information on thermal processing in the physical environment where the absorption occurs.  Whittet et al. (1988) showed that a relation exists between water-ice absorption optical depth ($\tau_{ice}$) and visual extinction toward young stars embedded within and observed through the Taurus star formation region.  Further work by Teixeira \& Emerson (1999) showed that the column density of water is a better tracer of the visual extinction toward obscured T Tauri and background stars.  

Table 4 presents parameters that characterize the water-ice absorption seen in the spectra.  Column 1 is the blackbody temperature that best fits the spectral continuum shape in the 2.0-2.3 and 3.5-4.0$\mu$m regions.  The additional values described in Table 4 were derived by dividing by the blackbody and removing the continuum shape from the spectra.  For IRAS 04181+2654B no blackbody temperature was found to fit the K and L-band continuum with less than 15\% average deviation, so a 2nd order polynomial fit was used.  The second and third columns in Table 4 are the wavelength of the peak absorption in the ice feature, $\lambda_{ice}$, and the FWHM of the water ice absorption, $\Delta\nu_{ice}$.  Columns 4 and 5 are the optical depth, $\tau_{ice}$, and the ice-band column density, N$_{ice}$ determined from the data, respectively.  Water ice column density is found using the relation N$_{ice}$ = $\tau_{ice}\Delta\nu$/A where $\tau_{ice}$ is the optical depth of the water ice, $\Delta\nu$ is the FWHM of the feature in cm$^{-1}$ and A is the integrated absorption cross section, A= $2.0\times10^{-19}$ molecules/cm$^2$ (d'Hendecourt \& Allamandola 1986).  

Uncertainties in the the FWHM of the water ice absorption can be large because of difficulty determining the water ice profile shape in the region of telluric absorption (2.55-2.9$\mu$m).  Young stars often have warm material in the inner regions of their circumstellar disks where water-ice has vaporized, so the 3$\mu$m absorption does not trace all material along the line of sight to the stars.    Because of these factors, the uncertainties in the visual extinction derived using the relations from Teixeira \& Emerson (1999a) or Whittet et al. (1988) can also be large.  As a result, relations between water-ice optical depth or column density and the visual extinction are not well calibrated for Class I protostars.

\subsection{2-4$\mu$m Spectral Synthesis Models}

To better determine the natures of the stars, a 2-step method for spectral synthesis modeling was used.  First, models were fit to the K-band spectra to determine the stellar type and infrared veiling at 2.22$\mu$m.  The resulting information was then used to generate a more complete model for visual extinction, ice absorption and infrared excess over the full 2-4$\mu$m range.  The underlying spectral type of the star is crucial for determining an accurate model, so fits were done only for the five stars that had photospheric absorption features characteristic to low mass stars.  IRAS 04264+2433 has modest infrared veiling and an estimated spectral type of M1$\pm$2 (Doppmann et al. 2005; White \& Hillenbrand 2004), but line ratio analysis (\S4) reveals that this star has a strong scattered light component even at infrared wavelengths.  I do not see infrared photospheric absorption features in this star, and since strong scattering would underestimate the extinction to the stellar photosphere I do not fit spectral models to it.  Models were also not generated for the other four stars that did not have spectroscopically determined stellar types.

\subsubsection{The K-band Model: Spectral Type and 2.22$\mu$m Veiling}

The K-band spectra for the 5 young stars with photospheric absorption features were used to determine the stellar spectral type, the infrared accretion excess (veiling) at 2.22$\mu$m, and an initial estimate of the visual extinction.  To model these spectra, I used a $\chi^2$ minimization code written in the IDL programming language which was kindly shared by L. Prato and is described in detail in Prato, Greene \& Simon (2003).   The routine was modified for use with the data obtained for this study.  The spectral type templates used for the analysis were from observations that I made for this purpose, as well as data from Cushing et al. (2003) that are available online.  

The young star spectrum was fit to a series of models of a spectral type reference star which had been veiled with an infrared emission excess and then reddened by extinction using an ISM extinction law  (A($\lambda$) = A$_v$[$0.55$/$\lambda$]$^{1.6}$; Prato , Greene \& Simon 2003).  The K-band veiling, r$_K$, is defined as the ratio of the infrared excess at 2.22$\mu$m to the intrinsic stellar continuum flux: F$_{IR}$(2.22$\mu$m)/F$_{int}$(2.22$\mu$m).  The excess is assumed to be wavelength independent.  A $\chi^2$ search for the infrared veiling value and a best-fit visual extinction was done for a given spectral type standard.  Typical models searched over values of r$_K$=0.0 to 1.5, and through a range of 20.0 mag in A$_v$ (typically from A$_v$=10.0 to 30.0).  The search was repeated for several different stellar spectral type templates, and the final model was adopted from the star which gave the best fit in the least squares sense to the continuum shape and photospheric absorption features of the YSO.  The wavelengths where H$_2$ and Br$\gamma$ emission is pronounced were not used in the range for the fit.

The spectral type and infrared veiling values at 2.22$\mu$m that best-fit the data are presented in the first two columns of Table 5.  Figure 2 shows the K-band spectra for each star, overplotted in blue is the best veiled and obscured spectral model.  The models give good to excellent fits to the continuum shape and photospheric absorption features of  Mg (2.28$\mu$m), Na (2.20$\mu$m), Ca (2.26$\mu$m) and the CO band-heads (2.29-2.38$\mu$m), even though the latter features can also be affected by non-photospheric absorption or emission in young stars.  In all cases, dwarf spectral type standards gave better fits than giant stars to the continuum and absorption features.  For IRAS 04181+2654A, 04181+2654B and 04295+2251 this analysis is the first spectroscopic determination of a stellar type for the underlying photospheres (though Doppmann et al. 2005 estimate a temperature of 3400K for 04295+2251).  For the other two stars, the spectral type determined here is consistent with previous estimates (White \& Hillenbrand 2004).

\subsubsection{The 2-4$\mu$m Spectral Model}

Once the best K-band models were generated, the spectral type and r$_K$s were used to generate more complete models over the full 2-4$\mu$m range.  The search over the full wavelength region allowed for a more accurate determination of the extinction and a fit of wavelength dependence to the excess emission.  Muzerolle et al. (2003) showed that infrared excess emission in the 2-5$\mu$m spectra of Classical T Tauri stars is consistent with blackbody emission with a single characteristic temperature.  The available laboratory data of water-ice absorbance also allows for fits of the 3$\mu$m profile shape to search for variations in ice temperatures in the circumstellar environment.

The synthesis model constructed for this test was adapted from the K-band spectral code to use a wavelength dependent infrared excess and an optical depth model which had ice profiles included.  The observed 2-4$\mu$m spectrum for each protostar is defined as:

F$_{obs}$= C$[($F$_{int}$+k($\lambda$))e$^{-\tau(\lambda)}$]

Where F$_{int}$ is the intrinsic photospheric flux of the young star estimated from the spectral standard.  k($\lambda$) is the wavelength dependent accretion excess, which is described using blackbody radiation with a single temperature, T$_{dust}$, constrained by the required r$_K$ value at 2.22$\mu$m determined by the K-band models (Muzerolle et al. 2003).  The wavelength dependent optical depth, $\tau(\lambda)$, has two components: one from the ISM extinction law as described for the K-band model in \S 3.1.1 (Prato, Greene \& Simon 2003), and a second optical depth component, $\tau_{ice}$, was derived directly from the laboratory absorbance data for water-ice (Gerakines et al 1995; 1996{\footnote{Laboratory water ice absorbance data are available online at: http://www.strw.leidenuniv.nl/$\sim$lab}}).  The absorption profiles seen in the stars were compared to water ice at 10K, 50K, 80K and 120K to search for thermal processing in the absorbing material.  The parameter C is a scaling constant to normalize the models to the flux for each star at 2.22$\mu$m in the calibrated spectra (Figure 1).  A $\chi^2$ minimization search was done over the three parameters, A$_v$, T$_{dust}$ and $\tau_{ice}$ for each of the stars in Table 5.

Early in the modeling process it became apparent that all of the targets exhibited the known absorption excess in the long wavelength wing of the 3$\mu$m feature (Van de Bult et al. 1985; Smith et al. 1989).   To date, it is still unclear what process or species causes this excess absorption.  Explanations for its existence could include nitrogen or carbon containing molecules frozen onto the water-ice mantles, affects of the profile shape from non-spherical grains and grain orientation, or overlaid absorption of a broad water vapor feature in the 3$\mu$m region (Smith et al. 1989; Zinov'eva 2005).  Regardless of the cause of the excess absorption, the success of the model fits required the 3.25 to 3.75$\mu$m wavelength region to be excluded in the $\chi^2$ analysis (consistent with the results of Smith et al. 1989).  The wavelengths with atomic and molecular hydrogen emission were also excluded in the fit.  The final models provided the best simultaneous fit to the young star spectra in the $\sim$2.0-2.4$\mu$m, 2.9-3.25$\mu$m and 3.75-4.04$\mu$m wavelength regions.

\subsubsection{Model Results}

The last four columns of Table 5 present values for A$_v$, r$_{L'}$, T$_{dust}$ and $\tau_{ice}$ from the best model for each target.  r$_{L'}$ is the L-band accretion excess at 3.76$\mu$m, defined as r$_{L'}$=F$_{IR}$(3.76$\mu$m)/F$_{int}$(3.76$\mu$m), and is a natural result of this spectral fit.  For all stars, L-band accretion excess values were 3.5 to 5 times greater than those derived for the K-band.  Figure 3 shows the 2-4$\mu$m spectra of the targets with the best model overplotted in red.   The visual extinctions from the best model were within $\sim$2.5 magnitudes of the values determined from the K-band model.  In all but one case, the models give a good fit to the K-band spectra, the 2.9-3.25$\mu$m wavelength region, and wavelengths greater than 3.8$\mu$m. 

The uncertainty in a given model was dependent predominantly upon A$_v$ and T$_{dust}$.  Changes in the continuum model shape did not strongly affect the fits for $\tau_{ice}$; variations of only $\sim\pm$0.07 in the optical depth were typical for a range of extinctions and dust temperatures.  For different normalization constants, several values for A$_v$ and T$_{dust}$ could fit the continuum shape (Aspin 2003).  However, uncertainties in A$_v$ and T$_{dust}$ can be quantified by investigating the reduced $\chi^2$ surface for a single value of C which normalized the best model to the flux calibrated young star spectrum (Prato, Greene \& Simon 2003).  Figure 4 presents the $\chi^2$ contour plot for the models to IRAS 04295+2251.  The contours of $\chi^2$=1.01, 2, 3, 6 and 10 are included.  Overplotted in blue in Figure 3e for IRAS 04295+2251 is a model that has a reduced $\chi^2$ value of 2.  The 1$\sigma$ uncertainties for the 2 parameter models are defined by the $\chi^2$=3.2 surface (approximated by $\chi^2$=3; Prato, Greene \& Simon 2003).  Figure 4 shows that the contours of the $\chi^2$ surfaces are narrower for A$_v$ than T$_{dust}$.  The models in the 2-4$\mu$m region were more dependent upon changes in extinction than dust temperature.  In most cases a narrow range of 2-3 A$_v$ was implied by the best fits.  

No model for IRAS 04181+2654B could be found which fit the K and L-band continuum shape simultaneously.  The ``best'' model presented in Figure 3c and plotted in red deviates from the spectral energy in the K-band continuum by $\sim$15\%, on average.  Also overplotted in the figure in green is the $\chi^2$ fit only to the K-band and 2.9-3.25$\mu$m region.  All models which fit the steep slope to the K-band continuum overestimated the longer wavelength flux by a factor of two.  The value for r$_K$ was 0.0 from the K-band model of IRAS 04181+2654B, so even in the absence of infrared veiling emission this object can not be well fit from attenuation by ISM extinction alone.  Discussion of the spectra and the nature of this strange source is continued in \S6.1.

The blackbody dust temperatures that fit the wavelength dependence of the infrared excess are in the range of $\sim$900-1050K.  This is warmer than the blackbody fits to the continuum in the raw data (Table 4), but not as high as the 1100-1400K temperatures derived for infrared excesses in the Classical T Tauris studied by Muzerolle et al. (2003).  Although infrared dust temperatures of greater than 1200K could not be excluded by the models, excesses with temperatures less than $\sim$650 were always ruled out with $\sim$3$\sigma$ confidence.  The visual extinctions determined for the spectral models are typically greater than values determined using other techniques (\S4).  The continuum shapes and deep profiles for the ice-band absorption are relatively well-fit by the models for IRAS 04181+2654A and 04295+2251.  However, the shallower absorptions seen in IRAS 04158+2805 and 04260+2642 are fit more poorly (see further discussion in \S6.1 for IRAS 04158+2805).   The ice absorption profiles were always better fit by optical depth models for cold ice (T$_{ice}$ $<$ 50K) versus warmer, annealed ice ($>$80K), indicating that significant thermal processing of the icy material along the line of sight toward these cool dwarfs has not occurred.

The 2-4$\mu$m modeling method provides a powerful way to determine the properties of young stars when the spectra can be well described by a veiled and reddened photosphere.  Yet, a significant fraction of the infrared flux from Class I protostars is scattered off the surface of the disk or envelope (Kenyon et al. 1993b; Whitney et al. 1997; Park \& Kenyon 2002).  The infrared images with high spatial sampling can still be dominated by scattered light (Padgett et al. 1999; Terebey et al. 2006).  Sellgren et al. (1996) showed that in the absence of PAH emission in the 3.3$\mu$m region, infrared reflection nebulosities can be approximated using Rayleigh scattered light with a 1/$\lambda^4$ wavelength dependence.  Ishii et al. (2004) modeled their spectra of an infrared reflection nebulosity near an embedded protostar with some success.  As a test, a very simple scattered light model with an isotropic phase function and Rayleigh wavelength dependence was introduced into the 2-4$\mu$m spectral fits.  This was accomplished by attenuating the standard plus infrared excess model, F$_{int}$+k($\lambda$), by Rayleigh scattered light before the star was reddened with ISM extinction.  This rudimentary investigation showed that if the contribution of scattered light is appreciable ($\sim$50\% of the flux at 2.2$\mu$m), the spectral models that do not take scattering into account will underestimate both the visual extinction and the infrared veiling.  As a result, the values for A$_v$ and r$_K$ determined from this analysis are probably lower limits for the zero scattered light case.

\section{Extinction Toward Embedded YSOs}

Accurate values for the visual extinctions are needed to derredden the observed fluxes and derive underlying stellar luminosities, but determining the obscuration toward an embedded protostar is challenging because of the unknown contribution from scattered light.  The available infrared photometry coupled with the 2-4$\mu$m spectral data provides five independent opportunities to measure visual extinctions toward these young stars (Table 6):

1) A$_v$(ice) - Determined by equating the column density of water ice along the line of sight to an A$_v$ using the relation presented in Teixeira \& Emerson (1999).  The applicability of the N$_{ice}$ to A$_v$ relation questionable; the scatter is large for embedded sources with A$_v>$20 determined from infrared photometry.  The column density will not trace warm circumstellar material where ices have sublimated off of dust grains (T$\ge$210K).  Hence, the extinction derived from water ice absorption is a very rough estimate of the amount of obscuring material along the line of sight to the embedded photosphere.

2) A$_v$(phot) - Estimated by dereddening J, H, and K-band magnitudes from 2MASS to the Classical T Tauri star locus (Meyer et al. 1997).

3) A$_v$(spec) -  Derived by using the 2-4$\mu$m spectral synthesis models to simultaneously determine infrared accretion excess and A$_v$ (see \S3.1).

4) A$_v$(H$_2$) - Found using molecular hydrogen line emission from the {\it v}=1-0 Q(3) and S(1) transitions which arise from the same state and have an intrinsic ratio of Q(3)/S(1)=0.7 (Turner 1977).  A$_v$ values are determined by assuming an ISM extinction law (defined in \S 3.1.1) and calculating the attenuation required to redden the 0.7 ratio to the observed value.  This line analysis provides a robust tracer of material along the line of sight to the emitting region, but the molecular hydrogen may not sample material toward the photosphere because it could arise from shocks that are spatially extended from the star.

5) A$_v$(Br$\alpha$/Br$\gamma$) -  Estimated by using Br$\alpha$ to Br$\gamma$ emission line ratios.  In a Case B treatment where Ly$\alpha$ photons are optically thick (Baker \& Menzel 1938), and for a wide range of temperatures and densities (assumed to be applicable to the atomic hydrogen emitting environments of young stellar magnetospheres), Br$\alpha$ is intrinsically 2.85 to 3.17 times brighter than Br$\gamma$ (Brockelhurst 1971; Hummer \& Storey 1987).  A$_v$s are estimated by assuming an intrinsic ratio of $\sim$3 and reddening this to the observed value using an ISM extinction law (\S 3.1.1). 

Table 6 summarizes the visual extinctions for the ten YSOs that were found using these different methods.  Each relies on assumptions that scattered light does not affect the visual extinction, and that the obscuration along the line of sight follows an ISM extinction law (as defined in \S 3.1.1).  Four of the stars have Q(3) and S(1) emission for H$_2$ line ratio analysis, five have spectral model fits, and seven have A$_v$s from Brackett ratios.  The independent extinction indicators give different estimates for A$_v$.  IRAS 04295+2251 is the only star with extinctions determined using all methods, and the resulting values differ by more than 20 mag.  Dereddening the infrared photometry is the most widely used technique in the literature.

Extrinsic to the stellar photosphere and circumstellar emitting region around a YSO, the infrared flux is attenuated by both extinction and scattering.  Emission line ratios can provide information on the decrease of flux from scattering compared to reddening.  Plotted in Figures 5 and 6 are the H$_2$ Q(3) versus S(1) line fluxes and the Br$\alpha$ versus Br$\gamma$, respectively.  In Figure 5, a line showing the intrinsic Q(3)/S(1) value is plotted, as are lines that show the behavior of the ratio when the emission is viewed through A$_v$=10mag and 20mag.  For the Brackett emission (Figure 6), similar lines are overplotted but a range of values instead of a single line is identified to take into account the ratios for differing physical environments of the atomic hydrogen emitting region (Brockelhurst 1971; Hummer \& Storey 1987).  For intrinsic hydrogen line emissions seen in pure (Rayleigh) scattered light, the H$_2$ Q(3)/S(1) ratio becomes 0.41, and the average Br$\alpha$/Br$\gamma$ ratio is $\sim$0.24.  Lines with these slopes are also overplotted in Figures 5 and 6.  

Data points to the upper left of the intrinsic line ratios in Figures 5 and 6 correspond to stars that have extinction dominating the scattered light for the ratio, and objects that lie to the lower right of the intrinsic line ratio are dominated by scattered light.  A combination of reddened and scattered light will make a star appear less obscured in Figures 5 and 6.  Most of the sources lie within the range of A$_v$ 0-15 magnitudes in the figures, even though higher extinctions of $>$20 mag are assumed to obscure Class Is and describe their SED shapes (Whitney et al. 1997).   IRAS 04108+2803B is the only star that has a visual extinction determined from the atomic hydrogen line ratio that is greater than for other methods.   

IRAS 04264+2433 has a molecular hydrogen line ratio of $\sim$0.4, which is less than the intrinsic ratio and is consistent with emission viewed in scattered light (Figure 5).  This source has no detectable Br$\alpha$ emission, but its Br$\gamma$ flux and Br$\alpha$ detection limit also put it underneath the intrinsic line ratio range and in the scattering dominated region of Figure 6.  This suggests that this star is viewed entirely in scattered light, which is consistent with the edge-on geometry adopted to explain observations at optical and mid-infrared wavelengths (see also \S6.1; White \& Hillenbrand 2004; Kessler-Silacci et al. 2005).

The derivation of multiple A$_v$ values (Table 6) was done to test the applicability of techniques used for Class II stars to the more obscured Class I and flat spectrum objects.  None of the extinctions presented in Table 6 can be deemed correct with a high level of confidence because of systematic affects such as infrared excess from disk emission (e.g., J, H, K photometry) and unknown levels of scattered light (Figure 5, 6).  Several authors are fitting the SEDs of young stars in conjunction with the scattered light images and polarization fractions  (Whitney et al. 1997; 2003a; 2003b; Eisner et al. 2005; Terebey et al. 2006).  This method for determining a valid A$_v$ value is promising.  The models of Whitney et al. (1997) predict that the attenuation of photospheric flux toward an embedded protostar can correspond to 20-60 or more magnitudes of optical obscuration.  The A$_v$ values derived here are at the lower end of this range.  In the future, detailed modeling of the multi-wavelength SEDs, coupled with polarimetric data and scattered light images, may make accurate visual extinctions available for the stars studied here (i.e., Whitney et al. 1997; 2003a; 2003b; Eisner et al. 2005).

\section{Accretion Luminosity and Mass Accretion Rates}

Muzerolle et al. (1998) showed that the dereddened Br$\gamma$ fluxes from Classical T Tauri stars correlate with accretion luminosities determined from the hot flux excesses measured in the ultraviolet.  Accurately dereddening Br$\gamma$ fluxes toward these protostars allows for the estimation of their mass accretion luminosities.  For the five stars that have spectral model fits, the Br$\gamma$ emission is dereddened using the visual extinctions from the best model.  IRAS 04108+2803B, 04361+2547 and 04365+2535 have some of the deepest and most pronounced water-ice absorption features in this sample, which suggests strong attenuation of the photospheric flux.  The visual extinctions adopted for these stars are the highest values from Table 6.   For the remaining two targets, the Br$\gamma$ fluxes are dereddened using the visual extinction from the infrared photometry.  

The visual extinctions used for dereddening the Br$\gamma$ fluxes are listed in the first column of Table 7.   Table 7 also presents the dereddened Br$\gamma$ line flux values, the derived accretion luminosities using Muzerolle et al.'s (1998) relation, the bolometric luminosity (from Chen et al. 1995) and the ratio of the latter two.  Uncertainties in the accretion luminosities are dominated by the ambiguities in the dereddening process (\S 4).  For example, if the A$_v$ is underestimated by 5 or 10 magnitudes, which seems plausible based on the values presented in Table 6, then the accretion luminosity will be correspondingly underestimated by factors of 1.9 and 3.7, respectively.  Also included in the lower section of Table 7 are the accretion luminosities from the data discussed by Muzerolle et al (1998) and a comparison to the bolometric luminosity.  IRAS 04361+2547 and IRAS 04239+2436 overlap in our studies, and the accretion luminosities that I derive are 2.5 to 3 times greater than the values they present.  The sources were chosen for this study because they showed variability in infrared flux.  Changes of a factor of 2.4 in the observed Br$\gamma$ flux will cause the accretion luminosity to increase threefold (assuming extinction is a constant).  Flux variations of this level are not surprising for variable YSOs.  

IRAS 04239+2436 is the only star that has an L$_{acc}$/L$_{Bol}$ of $>$0.5, and three stars have L$_{acc}$/L$_{Bol}$ in the 0.2-0.3 range.   All but two of the stars are actively accreting material from their circumstellar environments, but with the exception of IRAS 04239+2436 the accretion luminosities are not a dominant fraction of the total L$_{Bol}$.  The stars observed by Muzerolle et al. (1998) that have strong bolometric luminosities typically have lower L$_{acc}$/L$_{Bol}$ fractions than the variable YSOs in this sample.  

To relate the accretion luminosity to the mass accretion rate, information on the underlying stellar mass is required.  The discussion in \S4.1 of Doppmann et al. (2005) convincingly outlines problems using infrared spectroscopy, photometry, and theoretical evolutionary models to place embedded protostars directly on a HR diagram to determine their mass.   The applicability of evolutionary models toward very young stars is uncertain.  Moreover, the derived visual extinctions are known with so little confidence that dereddening infrared fluxes to find stellar luminosities is also questionable.  Nevertheless, models predict that a young star evolves vertically in luminosity along the Hayashi track in the 0.5-1.0 million year old (Myo) age range (Siess et al. 2000).  Thus, I follow the analysis of White \& Hillenbrand (2004) and estimate a Class I stellar mass based on the location of the one million year isochrone at the position of the temperature of the star.  The temperatures scale of White \& Hillenbrand (2004) for stellar spectral types is adopted.  The results are shown in the second column in Table 8.  The two stars that overlap with the White \& Hillenbrand (2004) sample have the same mass estimate in this study (IRAS 04158+2905 is substellar). 

Gullbring et al. (1998) estimated that the liberated energy from the accretion is related to the mass accretion rate, \.M, by using a Virial Theorom treatment: L$_{acc}$= GM$_*$\.M/R$_*$$\times$(1/R$_*$-1/R$_{in}$), where R$_{in}$ is the disk inner radius and has a typical value of 3-5R$_*$.  Using the mass estimates in Table 8, the corresponding luminosity for a 1 Myo star from the Siess et al. (2000) models to estimate radii, and a R$_{in}$ value of 5R$_*$, I have derived the mass accretion rates for the five stars in Table 8.

The accretion rates presented in Table 8 are similar to the average mass accretion values of $\sim$10$^{-8}$M$_{\odot}$/yr for Class II T Tauri stars (Gullbring et al. 1998).  These rates are several orders of magnitude too low to build the mass of a sun-like star in a $\sim$few $\times10^5$ years and are not consistent with envelope collapse models (Kenyon et al. 1990; Kenyon et al. 1993;  Muzerolle et al. 1998; White \& Hillenbrand 2004).  It is yet unclear if the unexpectedly low accretion rates for Class I stars derived by optical and infrared studies are because the main mass building phase in a stars life occurs either earlier than the Class I stage or by eruptive increases in the mass accretion (e.g. FU Ori or EXor outbursts; Kenyon et al. 1990 Calvet et al. 2000; White \& Hillenbrand 2004).  Additional knowledge of the variability of embedded protostars would be useful to investigate this.

IRAS 04158+2805 and IRAS 04260+2642 have mass accretion values derived by White \& Hillenbrand (2004), these values are also presented in Table 8.  Although we used the same underlying assumptions for stellar masses and luminosities, the values for \.M that I determine are more than an order of magnitude different from the ones they present.   The methods for determining these values have proven reliable for Classical T Tauri stars (Muzerolle et al. 1998), so it is tempting to attribute differences in mass accretion rate to the variable nature of these stars.  However, for IRAS 04260+2642, the limit for the accretion rate that I derived is 2.5 orders of magnitude less than the detected value from White \& Hillenbrand (2004).  Since the peak accretion rate enhancement in FU Ori outbursts is estimated to be 2-4 orders of magnitude (Calvet et al. 2000), the decrease seen in 04260+2642 seems excessively large for stellar variability over the one year time frame that the data for our respective works was obtained.  In the absence of further information on accretion variations in embedded protostars, it seems more probable that the standard techniques used to determine mass accretion in different wavelength regimes in Class II stars do not translate well to studies of embedded Class I protostars or YSOs observed strongly in scattered light.  

\section{The Nature of the Targets}

Five of the targets included in this study had prior evidence for outflow activity from observations reported in the literature, and the other five were not known to have outflows traced by optical emission species, molecular hydrogen emission or CO diagnostics at submm wavelengths (Gomez et al. 1997; Bontemps et al.1996; Moriarty-Schieven et al. 1992; Kenyon et al. 1998; White \& Hillenbrand 2004).  Eight stars show significant H$_2$ emission in their spectra, and several exhibit emission in multiple transitions (Figure 1).  One of the stars that does not show appreciable molecular hydrogen emission, IRAS 04365+2535, is known to drive a molecular outflow (Bontemps et al. 1996) and has  H$_2$ {\it v}=1-0 S(1) emission detected at a $\sim$3$\sigma$ level of confidence by Ishii et al. 2004.  In this sample, nine of the ten stars have H$_2$ emission.  It is not possible to determine conclusively between an origin of the H$_2$ from shocks in an outflow versus UV or X-ray fluorescent emission.  However, the youth and low masses and bolometric luminosities of these stars (low UV flux) weigh more toward an outflow origin. 

The existing degeneracy in the spectral index between true Class Is and reddened Class IIs has led a number of authors to try to distinguish the evolutionary states of young stars using other methods.  In the near-infrared, Park \& Kenyon (2002) used the spatial extent of surrounding reflection nebulosity to differentiate Class Is from transition or reddened T Tauris.  Infrared accretion excess emission is usually strong in embedded protostars, and values of r$_K<$0.2 are rarely seen for Class Is (Greene \& Lada 1996; Luhman \& Rieke 1999; Doppmann et al. 2005).  Kessler-Siliacci et al. (2005) also proposed an evolutionary sequence based on the 10$\mu$m profile of the silicate absorption and emission; more embedded protostars exhibit strong absorption and the more evolved T Tauris exhibit only silicates in emission.  Though, the use of the silicate diagnostic alone is ambiguous because source geometry can also cause differences in 10$\mu$m profiles (Ghez et al. 1991; Chiang \& Goldreich 1999; Ciardi et al. 2005).  Additionally, Motte \& Andr\'e (2001) devised an evolutionary sequence based on envelope masses in which the younger protostars have resolved mm emission with M$_{Env}$ greater than $\sim$0.1 M$_{\odot}$, and more evolved sources have unresolved emission and envelope masses less than this.   Scattered light levels were determined by Whitney et al. (1997), and the embedded protostars were found to have higher infrared polarization fractions. 

Table 9 presents a summary of the observed properties of these ten young stars.  The first and second columns are the bolometric temperatures and spectral indices, which are two parameters that have been used in the past to define the Class Is (Chen et al 1995; White \& Hillenbrand 2004).  Also included are the L$_{acc}$/L$_{Bol}$ fractions and water-ice optical depths from this work (columns 4 and 5) and data collected from Kessler-Siliacci et al. (2005) and Watson et al. (2004) for the silicate emission (column 6), Motte \& Andr\'e (2001) for envelope masses (column 7), and Whitney et al. (1997) for polarization values (column 8).   

Figure 7 presents the spectra for the 10 stars plotted from bottom to top in order of increasing K-L$'$ color.  Included to the left of the figure is a note on the order of the spectra from Figure 1.  Seven YSOs have water-ice optical depths greater than one indicating a high level of obscuration along the lines of sight, but no apparent relation between $\tau_{ice}$ and other characteristics is seen.  However, trends between the other properties from Table 9 are found.  The stars that have 10$\mu$m silicates in emission typically have lower or unresolved envelope masses, lower L$_{acc}$/L$_{Bol}$ fractions, and low polarization fractions.  These stars are in transition between Class I and Class II or are reddened Class IIs.  Conversely, the embedded protostars have higher average L$_{acc}$/L$_{Bol}$ fractions, silicates in absorption (where data is available), higher envelope masses and stronger polarization fractions.  Although their spectral indices and bolometric temperatures are consistent with a Class I designation, IRAS 04108+2308B and IRAS 04264+2433 show many characteristics consistent with a more evolved nature.

Overlaid on the right side of Figure 7 is a note regarding the nature of these stars from the accumulated data (Table 9).  IRAS 04108+2308B is plotted as an embedded star in Figure 7 to avoid overlapping spectra, but this classification may be incorrect (see discussion below).  The results presented here are largely consistent with the classifications of these YSOs from the other studies.  However, I identify IRAS 04181+2654B and IRAS 04295+2251 as reddened Class IIs because they have unusually low infrared accretion veiling emission and several indicators of a more evolved nature (see also \S6.1). Observations show that 40-50\% of the stars identified as Class Is from their spectral indices are more evolved stars seen strongly in scattered or reddened light (Whitney et al. 1997; 2003; Motte \& Andr\'e 2001; Park \& Kenyon 2002;  White \& Hillenbrand 2004).  In this work, I find the sample of ten variable YSOs to be comprised of five true protostars, three in transition between Class I and II, and two reddened Classical T Tauris. 

\subsection{Notes on Individual Sources}

{\it IRAS 04108+2803B:}  Strong atomic hydrogen emission features (Br$\gamma$, Br$\alpha$, Pf $\gamma$) and molecular hydrogen {\it v}=1-0 S(1) emission are detected in IRAS 04108+2803B.  No absorption from photospheric features seen in young solar type stars is found, presumably because the infrared excess emission overwhelms the absorptions.  This object shows a deep and pronounced water-ice absorption feature. It has low envelope mass and unresolved mm emission (Motte \& Andr\'e 2001), low level of polarized infrared light (Whitney et al. 1997), and 10$\mu$m silicates detected in emission (Kessler-Siliacci et al. 2005).  Chiang \& Goldreich (1999) also fit the SED of this star as a CTTS viewed through obscuration in its inclined circumstellar disk.  04108+2308B is plotted as a Class I star in Figure 7 (to avoid overlapping spectra), but based on the multi-wavelength data listed in Table 9 this placement is probably erroneous. It seems more likely that IRAS 04108+2308B is strongly reddened but in transition between the Class I and Class II phase, or is an obscured Class II.

{\it IRAS 04158+2805:}  Weak atomic hydrogen emission features and molecular hydrogen emission are seen in IRAS 04158+2805.  The spectra show photospheric absorption features with a best fit spectral type of M6$\pm$1.  The infrared veiling value of r$_K$ of 0.25 derived for this star is significantly less than the r$_K$=1.1 determined by Doppmann et al. (2005), suggesting variable infrared accretion excess emission.

The 3$\mu$m water absorption feature observed in IRAS 04158+2805 is broad and shallow with no distinct rounded peak as seen in most other sources.   Particularly, the peak absorption is at 3.12$\mu$m for IRAS 04158+2805 (Table 4), which is shifted with respect to the absorption peak of the other stars.  The data were not well fit by the laboratory ice-band absorbance data (Figure 3a).  With a spectral type of M6, this source is potentially a Class I brown dwarf (White \& Hillenbrand 2004).  Late spectral type stars with cool photospheres have long been known to exhibit pronounced photospheric water vapor (Woolf 1964).  Indeed, the broad and poorly peaked absorption profile at 2.9-3.6$\mu$m and excess absorption in the 2.3-2.5$\mu$m region seen in IRAS 04158+2805 is more characteristic of vapor absorption in very late type Mira variables than it is of rounded or peaked ice profiles seen in embedded YSOs (Elias 1978; Whittet et al. 1988; Sato 1990).  It seems plausible that a pronounced photospheric water vapor feature is confusing the identification of water-ice absorption.  High spatial resolution images show that this source is probably dominated by scattered light at 2$\mu$m (Duchene et al. in prep), so the relative fraction of scattered versus transmissed light may also affect the continuum shape in the 2-4$\mu$m spectra.  The available data (Table 9) suggest that IRAS 04158+2805 is a late-type star in transition between Class I and II phases.  Though, White \& Hillenbrand (2004) pointed out that the low and unresolved envelope mass (Table 9) may be a consequence of the lower central stellar mass, so additional data is required to confirm this interpretation.

{\it IRAS 04181+2654A:}  This star shows weak photospheric absorption features, and a best spectral type of M3$^{+2}_{-1}$ is found.  The K-band accretion excess of r$_K$=1.3 is the largest veiling value derived from these data.  Atomic and molecular hydrogen emission features are detected, and this star shows a deep water-ice absorption feature. IRAS fluxes and envelope masses are spatially unresolved with IRAS 04181+2654B.  The multi-wavlength data shows that this star is a  Class I.

{\it IRAS 04181+2654B:}  IRAS 04181+2654B is unique among the spectra in the sample; it is the only star that has neither molecular nor atomic hydrogen emission features indicating accretion or outflow activity.  The K-band region shows a very steep spectral shape with photospheric absorption features and a best-fit spectral type of K7$\pm$1.   Photospheric absorption is often veiled to undetectability by infrared disk emission in young protostars.  Yet, in contrast to the spectra of 04181A, B shows no evidence for infrared veiling emission.  Additionally, IRAS 04181+2654B is undetected in the optical, the estimated I-band magnitude is $>$24, consistent with the very large visual extinction derived from the infrared photometry and the K-band spectral modeling (\S3).  The extinction derived from the K-band spectra corresponds within 1$\sigma$ the extinction determined from the 2MASS photometry.  This result indicates that there is not a significant amount of scattered light or infrared accretion excess emission, both of which affect the A$_v$ determinations from the photometry of Class I protostars.

IRAS 04181+2654B shows none of the near-infrared spectral indicators of youth, which argues against a Class I nature.  Indeed, its protostellar designation is based on spatially unresolved IRAS fluxes with 04181A.  IRAS 04181+2654B is too bright in the infrared to simply be a background K dwarf or giant star reddened by 40 magnitudes of extinction.  A K dwarf observed behind the Taurus dark clouds through $\sim$40 mag of obscuration would have an estimated K magnitude of $\sim$15, which is 4 mag fainter than 04181B appears (Table 1).  Moreover, its 10$\mu$m spectrum has been modeled to have silicates in emission, which is an indicator of youth (Kessler-Silacci et al. 2005).  Curiously, the L-band flux is 50\% fainter than expected from a reddened, unveiled young star photosphere (Figure 3c).   The enhanced extinction and $\sim$35$''$ separation from 04181+2654A lead to the hypothesis that 04181+2654B could be a young star observed through the extended protostellar envelope of A (Motte \& Andr\'e 2001).  If this is the case, the assumption of ISM obscuration may be incorrect and this may be why the 2-4$\mu$m spectral models could not fit the continuum shape.  Further spatially resolved observations, particularly at mid-infrared wavelengths, are necessary to determine the true nature of this interesting star.  It seems most likely that 04181+2654B is misclassified as a Class I and is in fact a reddened Class II.

{\it IRAS 04239+2436:}  This is the most actively accreting star in this sample. The derived accretion luminosity is greater than $\sim$50\% of the total bolometric luminosity.  It shows emission in H$_2$ {\it v}=1-0 S(1) and a rich spectrum of atomic hydrogen emission from the Br$\delta$, Br$\gamma$, Br$\alpha$, Pf$\gamma$, Pf$\epsilon$, and the Hu 14, 15 and 17 transitions.  It also exhibits He I (2.06 $\mu$m) emission, the CO $\Delta$ {\it v}=2 band-heads in emission (also seen by Greene \& Lada 1996), and evidence for weak emission ($\sim$3$\sigma$) in the Na features at 2.206/2.208$\mu$m.  Presence of the Na features in emission imply that this star may be more massive than a typical K-M spectral type Class I star.  Additionally, He I emission is believed to arise from accretion driven winds in young stars and its presence here confirms a strong level of mass accretion (Edwards et al. 2003).  The spectra shows deep 3 $\mu$m water-ice absorption and a weak C-H stretch absorption at 3.47$\mu$m.  IRAS 04239+2436 is an actively accreting Class I star.

{\it IRAS 04260+2642:}  Marginal detection of atomic hydrogen emission features, yet this star has molecular hydrogen emission from six transitions in the 1.95 - 2.45$\mu$m spectral region.  It also shows photospheric absorption features characteristic to young stars and a spectral type of K5$\pm2$.   IRAS 04260+2642 has a low envelope mass and mm emission which is spatially unresolved (Motte \& Andr\'e 2001).  The optical forbidden emission lines from this star are narrow, and an edge-on orientation where velocities from the outflow are in the tangential direction was adopted to explain this (White \& Hillenbrand 2004).  IRAS 04260+2642 is likely in transition between the Class I and Class II phase.

{\it IRAS 04264+2433:} This star shows features from seven different H$_2$ transitions in the 1.95-2.45$\mu$m spectra.  The molecular hydrogen {\it v}=1-0 Q(3)/S(1) line ratio is less than the intrinsic value (Figure 6), which I interpret as being caused by strong scattered light.  IRAS 04264+2433 is the only star observed for this study that has a 10$\mu$m spectrum with silicates in emission only (Kessler-Siliacci et al. 2005).  The edge-on disk geometry adopted by previous studies (White \& Hillenbrand 2004; Kessler-Siliacci et al. 2005) is consistent with the data presented here.  This star is in transition between Class I/Class II and is seen very strongly in scattered light at optical and infrared wavelengths.   Prior to July 2005, IRAS 04264+2433 was listed incorrectly in the SIMBAD online database as Elias 6 (Park \& Kenyon 2002, Motte \& Andr\'e 2001).  

{\it IRAS 04295+2251:}  This star exhibits deep water-ice absorption, atomic and molecular hydrogen emission, and absorption features typical of the photospheres of young low mass type stars with a best spectral type of K7$\pm$2.  IRAS 04295+2251 has the lowest infrared excess emission detected in this study, r$_K$=0.09, and a low veiling value was also determined by Doppmann et al. (2005).  The small infrared excess coupled with the fact that the 10$\mu$m spectrum has silicates in emission, the infrared polarization is low, and the star has an unresolved and low mass envelope leads to the proposition that this is a strong candidate reddened Class II star.  Eisner et al. (2005) found that the SED and optical/mm images of this star are best fit with a disk+obscuration model, although it could not be discerned if excess obscuration lies in an extended envelope (Class I star) or could be from ambient cloud material (reddened Class II).   Density enhancements of $\sim$20 mag are possible through the Taurus star forming region  to redden IRAS 04295+2251 if it was located behind the clouds (Whittet et al. 1988; Teixeira \& Emerson 1999).  Yet, unlike for IRAS 04181+2654B, this young star is revealed in the i-band (White \& Hillenbrand 2004; Eisner et al. 2005).  The strong visual extinction of A$_v$=28.0 mag determined for IRAS 04295+2251 from the 2-4$\mu$m spectral model must be caused in part by local disk or envelope material which attenuates the infrared light but scatters the optical flux (Eisner et al. 2005).  Based on the discussion presented here and values listed in Table 9, IRAS 04295+2251 may be a reddened Class II star.

{\it IRAS 04361+2547:}  This object star several atomic hydrogen emission features and has a rich molecular hydrogen emission spectrum from seven transitions at 1.95-2.45$\mu$m.    It also exhibits deep water ice absorption and absorption in the C-H stretch mode oscillation at $\sim$3.47$\mu$m.  IRAS 04361+2547 also has the highest envelope mass and greatest level of infrared polarization of any star observed for this study. There is no evidence for photospheric absorption features.  IRAS 04361+2547 is a true Class I protoster. 

{\it IRAS 04365+2535}:  Strong atomic hydrogen emission features but marginal molecular hydrogen emission ($\sim$2$\sigma$ at 2.12$\mu$m) are seen toward this source.  It also shows continuum flux that rises steeply toward longer wavelengths, a very deep, rounded 3 $\mu$m water ice absorption feature, and absorption in the C-H stretch mode oscillation at $\sim$3.47 $\mu$m.  There is no evidence for photospheric absorption features and high envelope mass and strong polarization confirm its nature as a true Class I embedded protostar.   





\section{Summary}

1)  New photometry and 2-4$\mu$m spectra are presented for ten Class I and flat spectrum young stars previously found to have variable infrared flux.  Molecular and/or atomic hydrogen emission features are detected in all except for IRAS 04181+2654B, and five stars show photospheric absorption features characteristic to low mass stars.  Evidence for water absorption at 3$\mu$m is seen in all but IRAS 04264+2433.

2)  For the five stars that show photospheric absorptions, a two step process was carried out to fit models to the 2-4$\mu$m spectra.  Spectral types, ice absorption characteristics, visual extinctions and blackbody fits to the infrared accretion excess were done.  For three of the stars, the model presents the first spectral type determination, and the other two are consistent with past studies.  The stellar spectra were best fit by models with cooler accretion excess dust temperatures than Classical T Tauris, visual extinctions that were typically higher than found using other methods, and water-ice absorption profiles characteristic to cold dust grains.

3)  Visual extinctions determined for the young stars using five techniques gave very different results.  The commonly used method of dereddening infrared fluxes gave A$_v$ values in the middle range compared to other techniques, and higher values for the extinctions were typically found from the 2-4$\mu$m spectral model fits.  All of the derivations of extinction assumed that the obscuring material follows an ISM law, and that scattered light is not an important factor.  This analysis was done to demonstrate the wide range of A$_v$ values obtained by applying techniques used for CTTSs to the more obscured Class I and flat spectrum protostars.

4) Extinctions determined by dereddening atomic and molecular hydrogen line ratios resulted in low values (0-15mag) for most stars, and negative values two cases.  This suggests that scattered light does strongly affect infrared fluxes.   The line ratios for IRAS 04264+2433 are consistent with an orientation where the infrared flux is seen entirely in scattered light  (such as a disk viewed edge-on).  Determining a true value for the photospheric attenuation should take scattering into account, and simultaneous modeling of the SED with scattered light images or polarimetry provide the most promising technique (Whitney et al. 1997; 2003; Eisner et al. 2005; Terebey et al. 2006).

5)  Br$\gamma$ line fluxes were dereddened using a best estimate for the extinction, and accretion luminosities (and limits) were determined for all stars.  As from past works, the stellar accretion luminosities are not as high as predicted from envelope collapse models.  A difference in \.M of 2.5 orders of magnitude is found for IRAS 04260+2642 from this study and results from optical spectroscopy (White \& Hillenbrand 2004).  This seems excessive for variable mass accretion for a one-year timescale (\S5), and suggests instead that the techniques used to derive \.M at different wavelengths for Classical T Tauri stars may have systematic problems when applied to younger stars.  

6)  Characteristics of the 2-4$\mu$m spectra of the ten stars are placed in the context of existing data from the literature.  Trends are seen between the observed multi-wavelength properties; stars with high envelope masses and polarization fractions typically have higher accretion luminosities and silicates in absorption, whereas stars with low, unresolved envelope masses and low polarizations have evidence for silicates in emission and lower levels of accretion.  From the accumulated data and new information presented here, two of the stars are best classified as reddened T Tauris, and three are stars transition between Class I and II and seen strongly in scattered light.

\acknowledgments

I am grateful to Lisa Prato and Saeid Zoonematkermani for sharing their spectral fitting code for determining stellar spectral types, veiling values and visual extinction from K-band spectra of young stars.  I thank Lisa Prato and Michal Simon for providing valuble suggestions during the course of this project, and Colin Aspin for reading the manuscript and offering advice for its improvement.  I also thank the anonymous referee who provided comments that improved the paper.  I appreciate the support provided by the IRTF staff - Bill Golish and Bobby Bus for their help with observing, and John Rayner and Mike Cushing for advice on reduction of SpeX data.   This publication makes use of data products from the Two Micron All Sky Survey, which is a joint project of the University of Massachusetts and the Infrared Processing and Analysis Center/California Institute of Technology, funded by the National Aeronautics and Space Administration and the National Science Foundation.  This research has made use of the SIMBAD database, operated at CDS, Strasbourg, France, and was supported by the Gemini Observatory, which is operated by the Association of Universities for Research in Astronomy, Inc., on behalf of the international Gemini partnership of Argentina, Australia, Brazil, Canada, Chile, the United Kingdom, and the United States of America.

\begin{deluxetable}{lccccccc}
\tabletypesize{\scriptsize}
\tablecaption{Target List and New Photometry \label{tbl-1}}
\tablewidth{0pt}
\tablehead{
\colhead{Star Name (IRAS \#)} & \colhead{RA J(2000)}   & \colhead{Dec. J(2000)}  &  \colhead{SED Class}  & \colhead{$\sigma_K^1$} & \colhead{K mag} & \colhead{L$'$ mag}}
\startdata
04108+2803A & 04 13 53.26  & +28 11 23.6  & Class I & 0.04  & 10.32 & 9.4 \\
04108+2803B & 04 13 54.69  & +28 11 33.1  & Class I & 0.52   & 10.65 & 7.9  \\
04158+2805  & 04 18 57.76  & +28 12 22.7 & Class I & 0.16  & 10.66 & 9.4 \\
04181+2654A & 04 21 11.47  &  +27 01 09.4 & Flat & 0.24  & 10.47 & 8.5   \\
04181+2654B  & 04 21 10.39  & +27 01 37.3 & Flat & 0.19   & 10.98 & 9.4  \\
04239+2436  & 04 26 57.1 & +24 43 36 & Class I & 0.31  & 9.88 & 7.2 \\
04260+2642  & 04 29 05.03 & +26 49 04.4 & Flat & 0.27  & 11.33 & 10.3  \\
04264+2433  & 04 29 30.08 & +24 39 55.1 &  Class I & 0.18 & 11.27 & 10.7  \\
04295+2251  & 04 32 32.07 & +22 57 30.3 & Flat & 0.15 & 10.05 & 8.7 \\
04361+2547  & 04 39 13.87 & +25 53 20.6   & Class I & 0.21  & 10.32 & 8.4 \\
04365+2535  & 04 39 35.01 & +25 41 45.5 &  Class I &  0.22 & 10.42 & 7.0 \\
\enddata
\tablenotetext{1}{$\sigma_K$ is the standard deviation of multiple K-band brightness measurements (in magnitudes) of these Class I stars, as determined in the study of Park \& Kenyon (2002).}
\end{deluxetable}

\clearpage


\clearpage

\begin{deluxetable}{lcc}
\tabletypesize{\scriptsize}
\tablecaption{Detected Spectral features \label{tbl-3}}
\tablewidth{0pt}
\tablehead{
\colhead{Species} & \colhead{Wavelength ($\mu$m)}   & \colhead{Note}}
\startdata
H$_2$ S(0) & 1.97 & 2 \\
H$_2$ S(2) & 2.03 & 2 \\
He I & 2.06 & 6 \\
H$_2$ S(1) & 2.12 & 2 \\
HI Brackett $\gamma$ & 2.16 & 1 \\
Na         & 2.20 & 3 \\
H$_2$ S(3) & 2.22 & 2 \\
Ca         & 2.26 & 3 \\
C0         & 2.29 & 3,4 \\
C0         & 2.32 & 3,5 \\
C0         & 2.35 & 3,4 \\
C0         & 2.38 & 3,4 \\
H$_2$ Q(1) & 2.40 & 2 \\
H$_2$ Q(2) & 2.41 & 2 \\
H$_2$ q(3) & 2.42 & 2 \\
H$_2$O ice & $\sim$3.05$\mu$m & 5 \\
C-H Stretch & $\sim$3.47$\mu$m & 5 \\
HI Pfund $\gamma$ & 3.74 & 1 \\
HI Brackett $\alpha$ & 4.05 & 1 \\
\enddata
\tablecomments{1 - Magnetospheric emission from material accreting onto the star, 2 - Molecular hydrogen emission, likely from shocks in outflows, 3 - Photospheric absorption features characteristic to young stars, 4 - CO band-head emission from warm inner regions of YSO accretion disks or winds, 5 - Broad, solid-state water ice absorption and shallow C-H stretch absorption from ices frozen onto dust grains along the line of sight. 6 - Helium I emission from accretion driven winds in stars with strong mass accretion. }
\end{deluxetable}

\clearpage
\begin{deluxetable}{lcccccccccc}
\tabletypesize{\scriptsize}
\tablecaption{Spectral Line Fluxes and Detection Limits ($\times$10$^{-18}$ W/m$^2$/s) \label{tbl-3}}
\tablewidth{0pt}
\tablehead{
\colhead{Star Name} & \colhead{Br$\gamma$}   & \colhead{Br$\alpha$}  & \colhead{Pf$\gamma$}    & \colhead{H$_2$ S(0)}  & \colhead{H$_2$ S(2)}  & \colhead{H$_2$S(1)}  & \colhead{H$_2$S(3)}  & \colhead{H$_2$Q(1)}  & \colhead{H$_2$Q(2)}  & \colhead{H$_2$Q(3)}}
\startdata
04108+2803B & 3.4$\pm$0.4 & 96$\pm$4  & 27.7$\pm$0.6  & $<$2.1 (3$\sigma$)  &  $<$3 (3$\sigma$) & 1.3$\pm$0.4 &  $<$2.4 (3$\sigma$) & $<$3 (3$\sigma$)  & $<$3 (3$\sigma$)  & $<$3 (3$\sigma$)   \\
04158+2805  & 1.8$\pm$0.6 & 10$\pm$2  &  $<$3.6  & $<$4.5   &  $<$3  & 2.7$\pm$0.5 &  $<$1.5  & $<$3   & $<$3   & $<$3    \\
04181+2654A  & 8.4$\pm$1.2 & 53$\pm$3  &  15$\pm$3    & $<$3   &  $<$3  & 3.5$\pm$0.9 &  $<$3  & $<$4.5   & $<$4.5   & $<$4.5    \\
04181+2654B  & $<$1.8  &  $<$6 &  $<$3    & $<$4.5   &  $<$3  & $<$2.4   &  $<$2.4  & $<$6   & $<$6   & $<$6    \\
04239+2436   & 53.2$\pm$0.6  & 319$\pm$4  &  117$\pm$2    &  $<$3   &  $<$2.4 & 5.7$\pm$0.6  &  $<$2.1  & $<$6   & $<$6   & $<$6    \\
04260+2642  & $<$2.1  & 4$\pm$1  &  $<$1.5  &  8.3$\pm$0.9   & 3.1$\pm$0.9 & 39.1$\pm$0.5 &  $<$2.1  & 11$\pm$2   & 4$\pm$2   & 8$\pm$2    \\
04264+2433  & 3.5$\pm$0.7  & $<$6  &  $<$4.5  &  22$\pm$1   & 4$\pm$1 & 9.6$\pm$0.5 &  4.5$\pm$1.5  & 12$\pm$1   & 6$\pm$1   & 9$\pm$1    \\
04295+2251 & 4$\pm$1 & 18$\pm$2 & $<$3 & 6$\pm$2 & 4$\pm$1 & 11$\pm$1 & 5$\pm$1 & 7$\pm$2  &  $<$4.5 & 4.5$\pm$1.5   \\
04361+2547 & 9.3$\pm$0.6 & 28.0$\pm$0.8 & 10$\pm$1 & 9.5$\pm$0.6 & $<$3 & 5.7$\pm$0.7 & 4$\pm$1 & 12.6$\pm$0.8  &  5.5$\pm$0.8 & 8.0$\pm$0.9   \\
04365+2535  & 13$\pm$1 & 131$\pm$2 & 27$\pm$3  &  $<$3 &  $<$2.1 &  2$\pm$1 &  $<$2.4 &  $<$3 & $<$3  & $<$3   \\

\enddata
\end{deluxetable}

\clearpage

\begin{deluxetable}{lccccc}
\tabletypesize{\scriptsize}
\tablecaption{Water Ice \label{tbl-4}}
\tablewidth{0pt}
\tablehead{
\colhead{Star Name} & \colhead{T$_{BB}$}   & \colhead{$\lambda_{ice}$}  & \colhead{$\Delta\nu_{ice}$}   &
\colhead{$\tau_{ice}$} & \colhead{N$_{ice}$}}
\startdata

04108+2803B & 640  & 3.05 & 550$\pm$60 &   1.49$\pm$0.05 &  4.0$\pm$0.5$\times10^{18}$ \\
04158+2805  & 910  &  3.12 &  910$\pm$90 &  0.6$\pm$0.1 &  2.6$\pm$0.8$\times10^{18}$ \\
04181+2654A & 750 & 3.04 &  780$\pm$90 & 1.26$\pm$0.07 &  4.9$\pm$0.6$\times10^{18}$  \\
04181+2654B  & --$^1$  & 3.08  & 700$\pm$100 & 2.2$\pm$0.2 &  8$\pm$2$\times10^{18}$ \\
04239+2436  & 670  &  3.05 &  510$\pm$90 & 1.5$\pm$0.1 &  4$\pm$1$\times10^{18}$ \\
04260+2642  & 980  & 3.06 &  630$\pm$90 & 0.39$\pm$0.09 &  1.2$\pm$0.5$\times10^{18}$ \\
04264+2433$^2$  & 1310 &  --- &  --- & --- & ---  \\
04295+2251  & 890 &  3.06 &  600$\pm$100 & 1.0$\pm$0.1 & 3.0$\pm$0.9$\times10^{18}$  \\
04361+2547  & 790  &  3.04 &  730$\pm$70 &  2.2$\pm$0.1 & 8$\pm$1$\times10^{18}$  \\
04365+2535  & 560 &   3.05 & 820$\pm$150 & 2.43$\pm$0.07 &  10$\pm$2$\times10^{18}$ \\
\enddata
\tablenotetext{1}{For IRAS 04181+2654B, no blackbody temperature could be found that had less than $\sim$15\% flux deviation from both the K and L-band continuum spectra.}
\tablenotetext{2}{The L-band spectra of IRAS 04264+2433 was too noisy to identify with confidence the optical depth and FWHM of a water ice absorption feature.}

\end{deluxetable}

\clearpage

\begin{deluxetable}{lcccccc}
\tabletypesize{\scriptsize}
\tablecaption{Spectral Synthesis Best-fit Models \label{tbl-1}}
\tablewidth{0pt}
\tablehead{
\colhead{Star Name} & \colhead{Spectral Type}   & \colhead{A$_v$}   &
\colhead{r$_K$ (2.22 $\mu$m)} & \colhead{r$_L$ (3.76 $\mu$m)} & \colhead{T$_{dust}$(K)} & \colhead{$\tau_{ice}$}}
\startdata
04158+2805  & M6$\pm$1 & 16.3  & 0.25$\pm$0.06 & 0.8$\pm$0.1 & 985 & 0.33  \\
04181+2654A  & M3$_{-1}^{+2}$ & 17.8 & 1.3$\pm$0.2 & 6.2$\pm$0.5 & 905 & 1.45 \\
04181+2654B  & K7$\pm$1 & 40.5 & 0.00$^{+0.04}_{-0.00}$ & -- & -- & --  \\
04260+2642  & K5$\pm$2 & 20.7 & 0.28$\pm$0.07  & 1.1$\pm$0.1 &  1020 & 0.38  \\
04295+2251  & K7$\pm$2 & 28.0 & 0.09$\pm$0.03 & 0.38$\pm$0.07 & 1040 & 1.12  \\
\hline
\enddata
\tablecomments{For IRAS 04181+2654B, the models that fit the steep slope of the K-band spectra had zero infrared veiling but always overestimate the L-band continuum shape by $\sim$50\% (See Figure 3c).}

\end{deluxetable}

\clearpage

\begin{deluxetable}{lccccc}
\tabletypesize{\scriptsize}
\tablecaption{Visual Extinction Determined Using Five Independent Methods \label{tbl-1}}
\tablewidth{0pt}
\tablehead{
\colhead{Star Name} & \colhead{A$_v$(ice)}   & \colhead{A$_v$(phot)}   &
\colhead{A$_v$(H$_2$)} & \colhead{A$_v$(spec)} & \colhead{A$_v$(Br$\alpha$/Br$\gamma$)}}
\startdata
04108+2803B & 19.1 & 16.3 & -- & -- & 33.9  \\
04158+2805  & 9.3 & 2.7 & -- & 16.3 & 9.2  \\
04181+2654A & 16.7 & 22.9 & -- & 17.8 & 12.1 \\
04181+2654B  & 26.6 & 40.8 & -- & 40.5 & --  \\
04239+2436  & 19.2 & 20.0 & -- & -- & 10.6  \\
04260+2642  & 7.3 & 6.0 & 15.7 & 20.7 & --  \\
04264+2433  & -- & 9.7 & neg. & -- & neg  \\
04295+2251  & 14.1 & 17.5 & 3.3 & 28.0 & 6.1  \\
04361+2547  & 26.4 & 20.8 & 9.1 & -- & 0.0  \\
04365+2535  & 29.3 & 12.0 & -- & -- & 18.4  \\
\enddata
\tablecomments{  ``Negative'' values of A$_v$ are derived for IRAS 04264+2433 from the H$_2$ and Br$\alpha$/Br$\gamma$ line ratios, which means that this star is viewed in the infrared predominantly in scattered light.}
\end{deluxetable}

\clearpage

\begin{deluxetable}{lccccc}
\tabletypesize{\scriptsize}
\tablecaption{Stellar Accretion Luminosity \label{tbl-1}}
\tablewidth{0pt}
\tablehead{
\colhead{Star Name} &  \colhead{Adopted A$_v$} & \colhead{Dereddened Br$\gamma$ Flux}   &
\colhead{L$_{acc}$ (L$_{\odot}$)} & \colhead{L$_{Bol}$ (L$_{\odot}$)$^1$} & \colhead{(L$_{acc}$/L$_{Bol}$)} }
\startdata 
\multicolumn{4}{c}{($\times$10$^{-18}$ W/m$^2$/$\mu$m)} \\
04108+2803B  & 33.9 & 112.0  & 0.15 & 0.6 & 0.25  \\
04158+2805  & 16.3 & 9.7 & 0.007 & 0.2 &  0.04 \\
04181+2654A & 17.8 & 52.7 &  0.06 & 0.3 &  0.2 \\
04181+2654B & 40.5 & $<$117 & $<$0.16 & 0.3 & $<$0.53 \\
04239+2436  & 20.0 & 386 &  0.7 &  1.3 &  0.54 \\
04260+2642  & 20.7 & $<$17.8 & $<$0.01 & -- & -- \\
04264+2433  & 9.7 & 9.5 &  0.007 &  0.3 &  0.02 \\
04295+2251  & 28.0 & 71.8 & 0.09 & 0.5 & 0.17 \\
04361+2547  & 26.4 & 142 & 0.20 & 2.5 &  0.08 \\
04365+2535  & 29.3 & 267 & 0.44 & 1.9 &  0.23 \\
\hline
\multicolumn{5}{c}{Additional Data from Muzerolle et al. (1998)}  \\
\hline
04361+2547  & -- & -- & 0.08 &  2.5 & 0.03 \\
04016+2610  & -- & -- & 0.03 &  3.6 & 0.008 \\
04239+2436   & -- & -- & 0.24 &  1.3 & 0.18 \\
04263+2426B  & -- & -- & 0.05 & 5.5 & 0.009 \\
04489+3042  & -- & -- & 0.02 &  0.3 & 0.07 \\
04292+2422  & -- & -- & 0.18 & 1.2 & 0.15 \\
\enddata
\tablecomments{$^1$ Bolometric luminosities are from Chen et al. (1995).}
\end{deluxetable}

\clearpage
\begin{deluxetable}{lccc}
\tabletypesize{\scriptsize}
\tablecaption{Stellar Mass Accretion \label{tbl-1}}
\tablewidth{0pt}
\tablehead{
\colhead{Star Name} & \colhead{1 Myo Mass (M$_{\odot}$)} & \colhead{\.M (M$_{\odot}$/yr)} & 
\colhead{log(\.M) (M$_{\odot}$/yr)}   }
\startdata
04158+2805  & 0.05 & 7.5$\times$10$^{-9}$ & -8.1   \\
04181+2654A  & 0.32 & 9.0$\times$10$^{-9}$ & -8.0 \\
04181+2654B  & 0.68 & $<$2.5$\times$10$^{-8}$ & $<$-7.6 \\
04260+2642  & 0.89 & $<$4.5$\times$10$^{-10}$ & $<$-9.3 \\
04295+2251  & 0.68 & 1.2$\times$10$^{-8}$ & -7.9 \\
\hline
\multicolumn{4}{c}{Additional Data from White \& Hillenbrand (2004)}  \\
\hline
04158+2805  & 0.05 & $<$3.2$\times$10$^{-10}$ & $<$-9.5   \\
04260+2642  & 0.89 & 1.6$\times$10$^{-7}$ & -6.79 \\
\enddata
\tablecomments{Approximate masses were derived by fitting the temperature estimate of the star to the 1 My isocrone using the Siess et al. (2000) evolutionary models (White \& Hillenbrand 2004).}
\end{deluxetable}

\clearpage
\begin{deluxetable}{l|cc|ccccc}
\tabletypesize{\scriptsize}
\tablecaption{Summary of Class I Properties \label{tbl-1}}
\tablewidth{0pt}
\tablehead{
\colhead{Star Name}  & \colhead{T$_{Bol}$} & \colhead{$\alpha$} &  \colhead{L$_{acc}$/L$_{Bol}$} & \colhead{ $\tau_{ice}$} & 
\colhead{Silicates} & \colhead{M$_{Env}^{4200AU}$ (M$_{\odot}$)} & \colhead{Polarization (P$_K$)}  } 
\startdata
04365+2535 & 172 & +1.27 & 0.23 & 2.43 & -- & $\sim$0.25 & 14.6\%  \\
04108+2803B & (205) & (+0.62) & 0.25 & 1.49 & Abs+Emis & 0.008U & 1.6\% \\
04239+2436 & 236 & +1.15 & 0.54 & 1.5  & Abs & 0.09 & 7.4\%  \\
04181+2654A & (346) & (+0.10) & 0.2 & 1.45 & Abs &  (0.12) & 1.9\% \\
04361+2547 & (144) & (+1.55) & 0.08 & 2.2 & --  & 0.25 & 21.2\% \\
04181+2654B  & (346) & (+0.10) & $<$0.55 & 2.2  & Emis+Abs & (0.12) & 1.7\% \\
04295+2251  & 447 & +0.21 & 0.21 & 1.12 & Emis+Abs & 0.025U & 2.6\% \\
04158+2805 & 528 & +0.72 & 0.04 & 0.33  & -- & 0.025U & -- \\
04260+2642  & -- & +0.25 & -- & 0.38 & -- & 0.025U & 1.8\%  \\
04264+2433  & (252) & (+0.98) & 0.02 & -- & Emis & 0.007U & 4.8\% \\
\enddata
\tablecomments{Values in parenthesis were derived from spatially unresolved fluxes (Chen et al. 1995; White \& Hillenbrand 2004; Motte \& Andr\'e 2001).  Accretion luminosities and water-ice optical depths are from this study, characteristics of the 10$\mu$m silicate profiles are from Kessler-Siliacci et al. (2005) and  Watson et al. (2004).  The Envelope masses within 4200AU are from Motte \& Andr\'e (2001), a U designates sources that were unresolved in their millimeter continuum study, Polarization values, P$_K$, are from Whitney et al. (1997). }
\end{deluxetable}

\clearpage






\clearpage

{\it Figure Captions:}

{\bf Figure 1:} The near infrared spectra of ten Class I and Transition protostars.  The upper panel for each star shows the full 2-4$\mu$m range, and the lower inset shows a zoomed view of the K-band region (1.96-2.45$\mu$m).  Emission and absorption features detected in the spectra are identified in the plots and described in Table 2.

{\bf Figure 2:} The K-band Spectra of five protostars that have photospheric absorption features.  Overplotted are best fit models derived by applying a (constant) veiling value to a spectral standard star and reddening it using ISM extinction.  The model results are presented in Table 5.

{\bf Figure 3:} The 2-4$\mu$m spectra of the 5 Class I protostars that have photospheric absorption features.  Overplotted are the best fit models derived by adding a wavelength dependent infrared excess emission to a spectral template standard, reddening spectrum using an interstellar extinction law (A($\lambda$)=A$_{v}$(0.55/$\lambda$)$^{1.6}$), and finally applying a wavelength dependent ice-band absorption model using available laboratory data for pure H$_2$O ice (See \S3).  The unexplained long wavelength wing of the water ice absorption feature is apparent in each of the spectra (3.25-3.8$\mu$m).  For IRAS 04181+2654B (panel c), the best model gave a poor fit to slope of the K-band. Overplotted in green is the best fit to the K-band spectra, extended to 4.0$\mu$m.  In panel e) for IRAS 04295+2251, overplotted in blue is a fit for a model that has a reduced $\chi^2=2$ residual and demonstrates uncertainties in individual model fits.

\clearpage

\begin{figure}
\plotfiddle{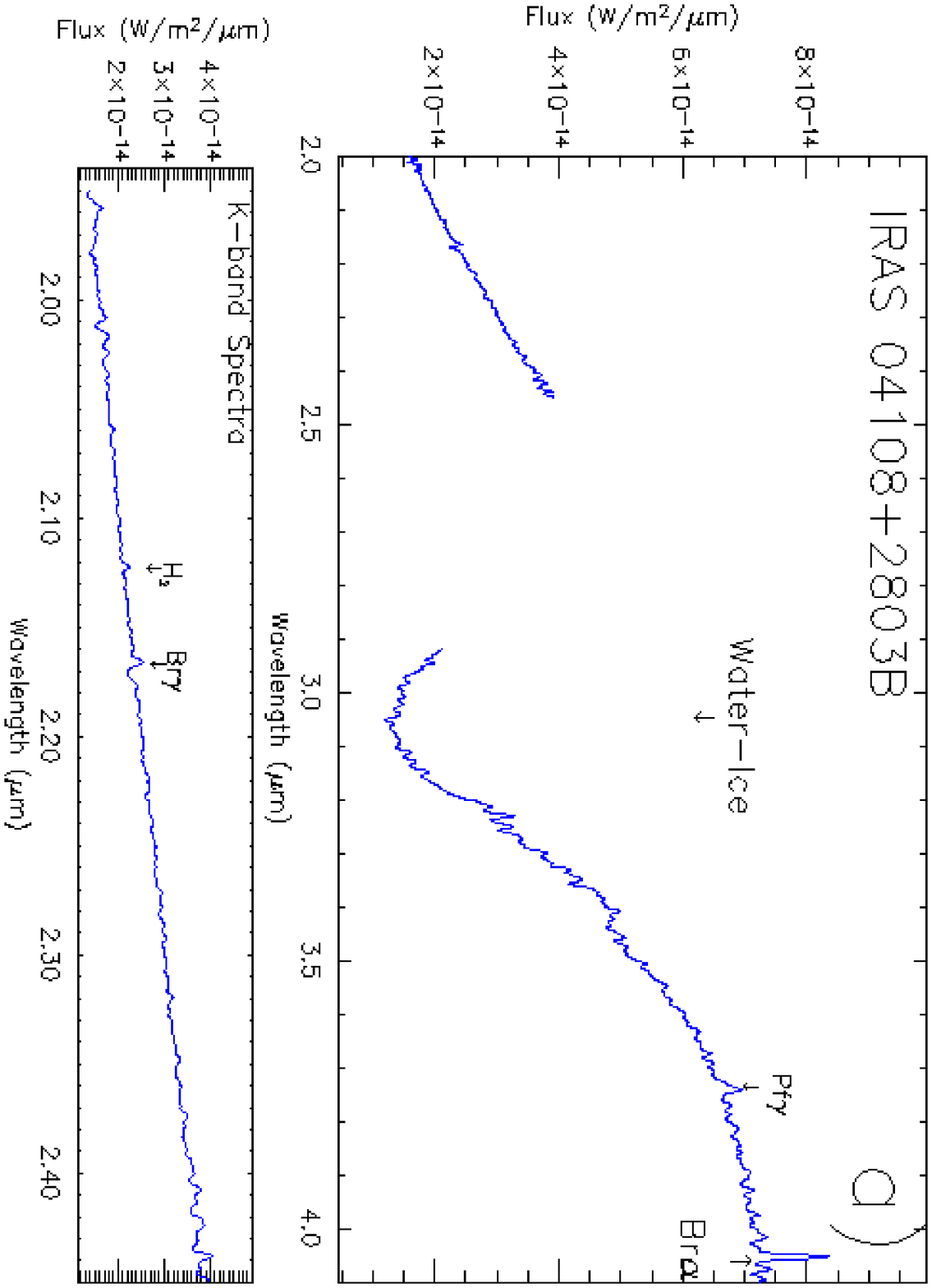}{350pt}{90pt}{40pt}{40pt}{10pt}{100pt}
\plotfiddle{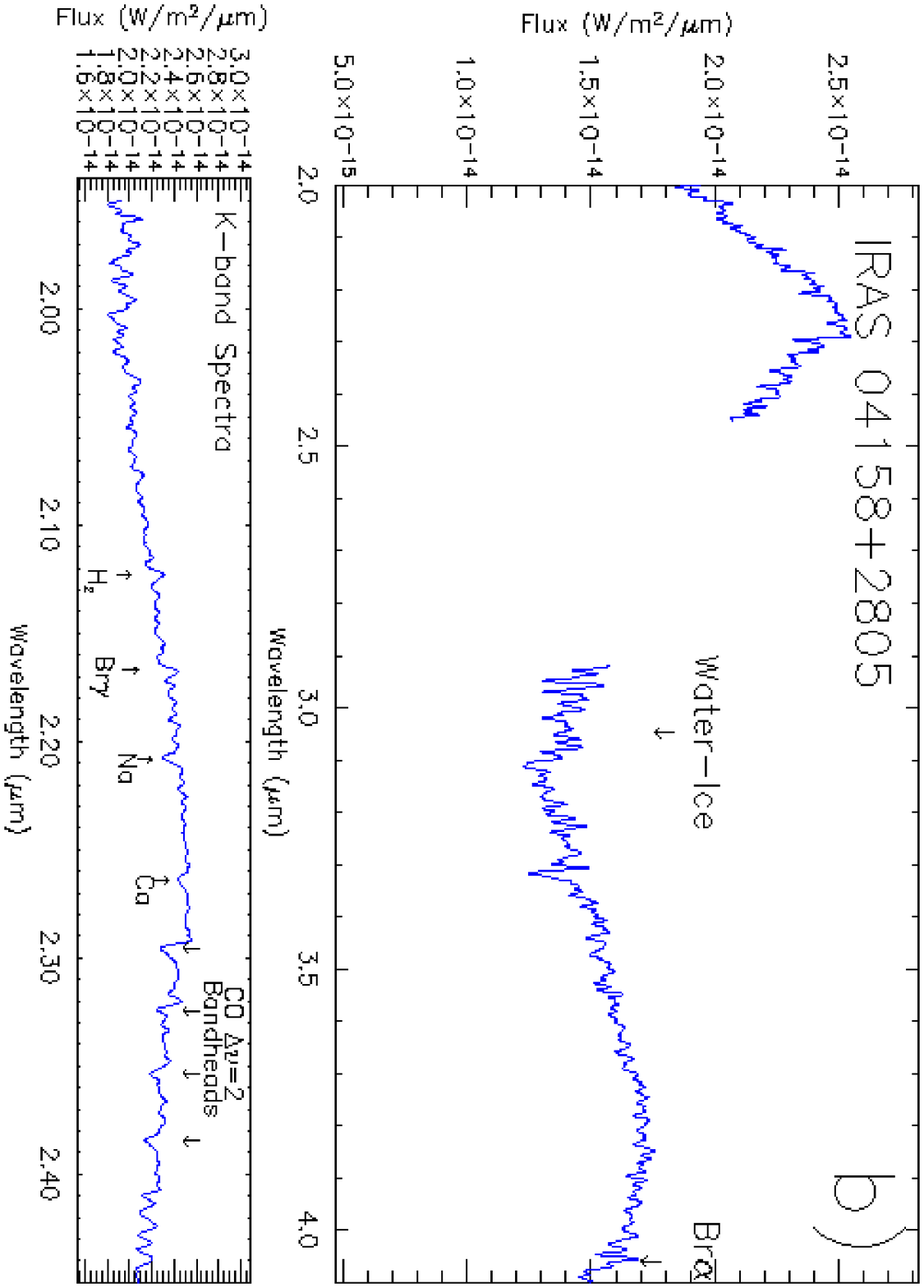}{350pt}{90pt}{40pt}{40pt}{290pt}{466pt}
\plotfiddle{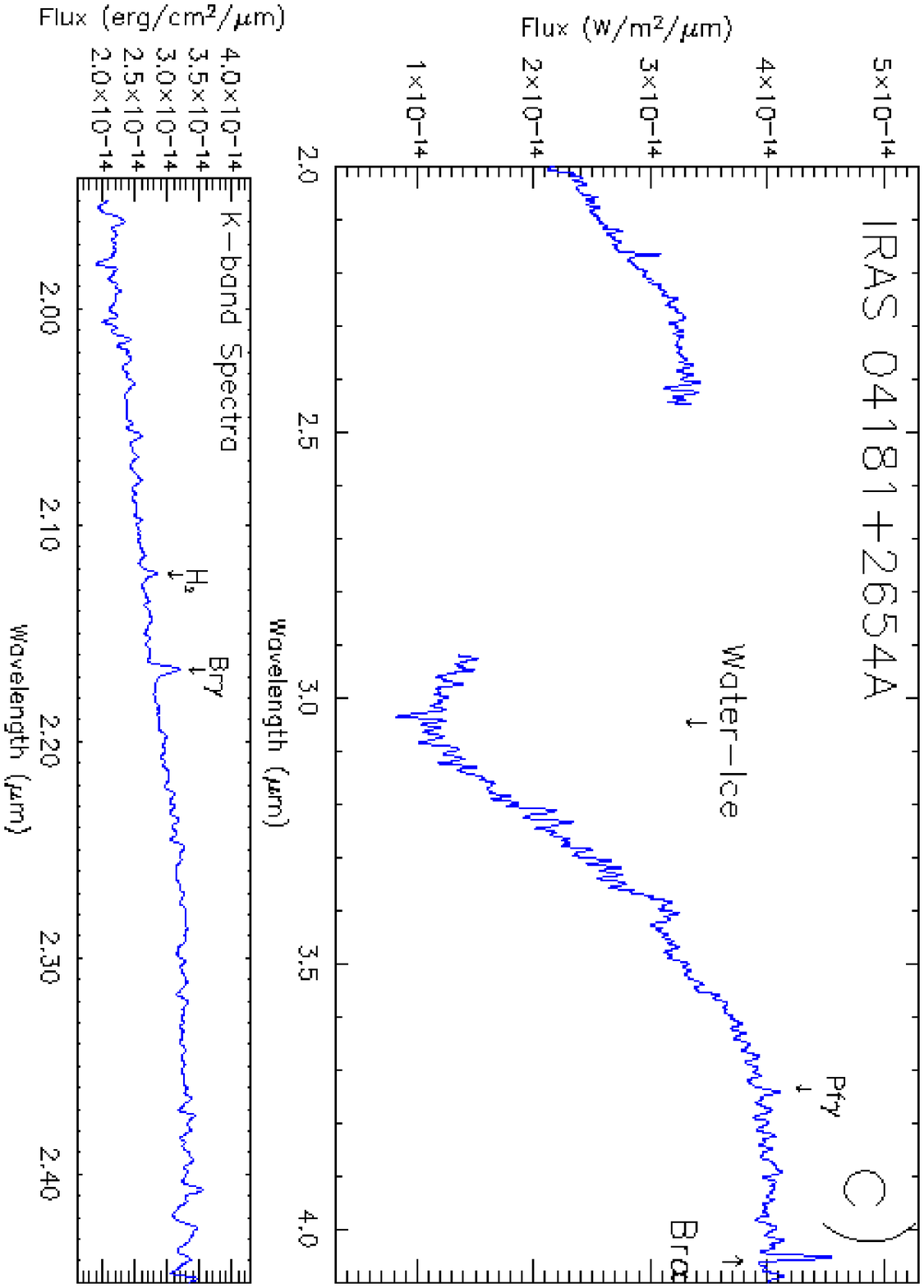}{350pt}{90pt}{40pt}{40pt}{10pt}{611pt}
\plotfiddle{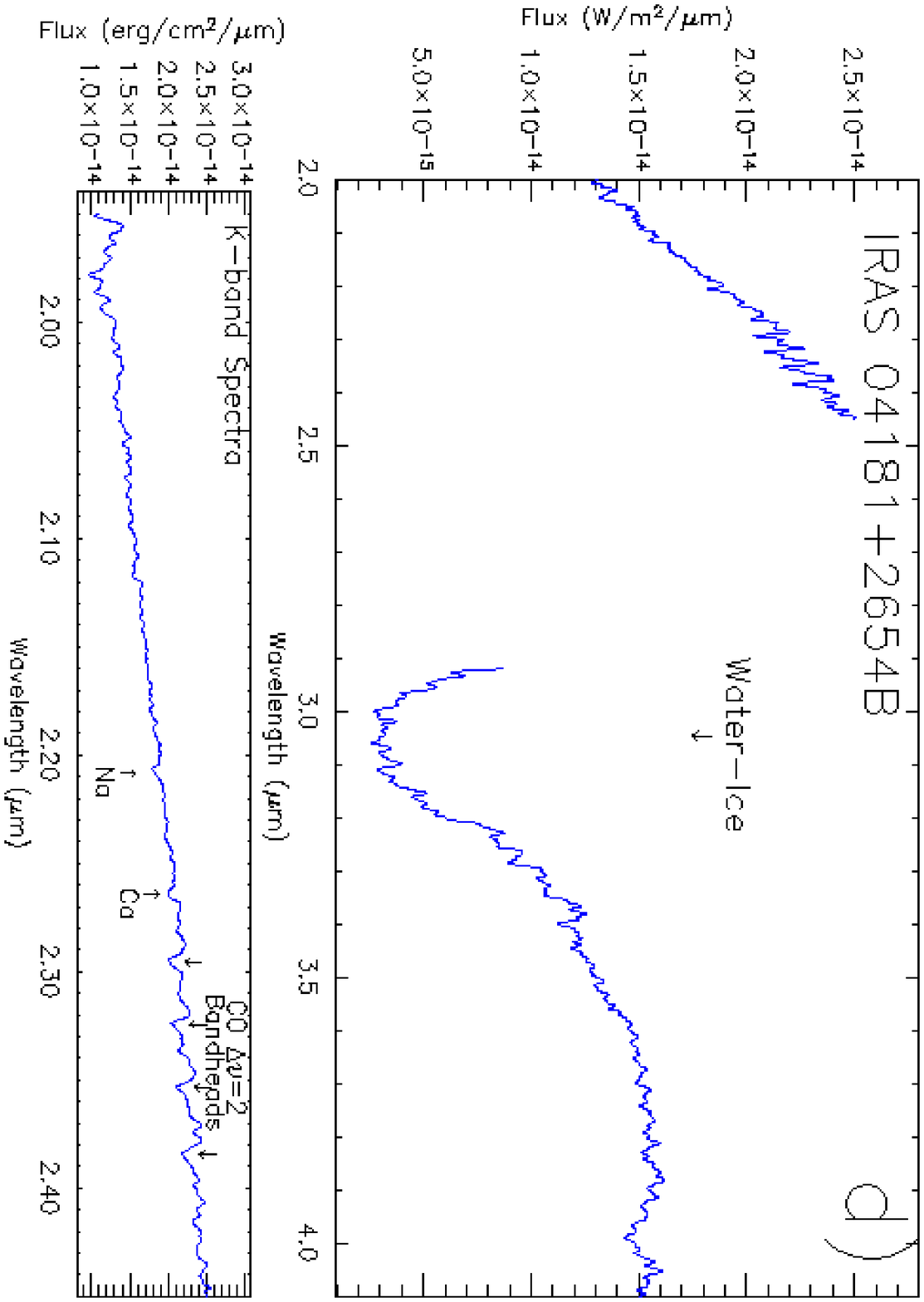}{350pt}{90pt}{40pt}{40pt}{290pt}{978pt}
\end{figure}
\newpage
\begin{figure}
\plotfiddle{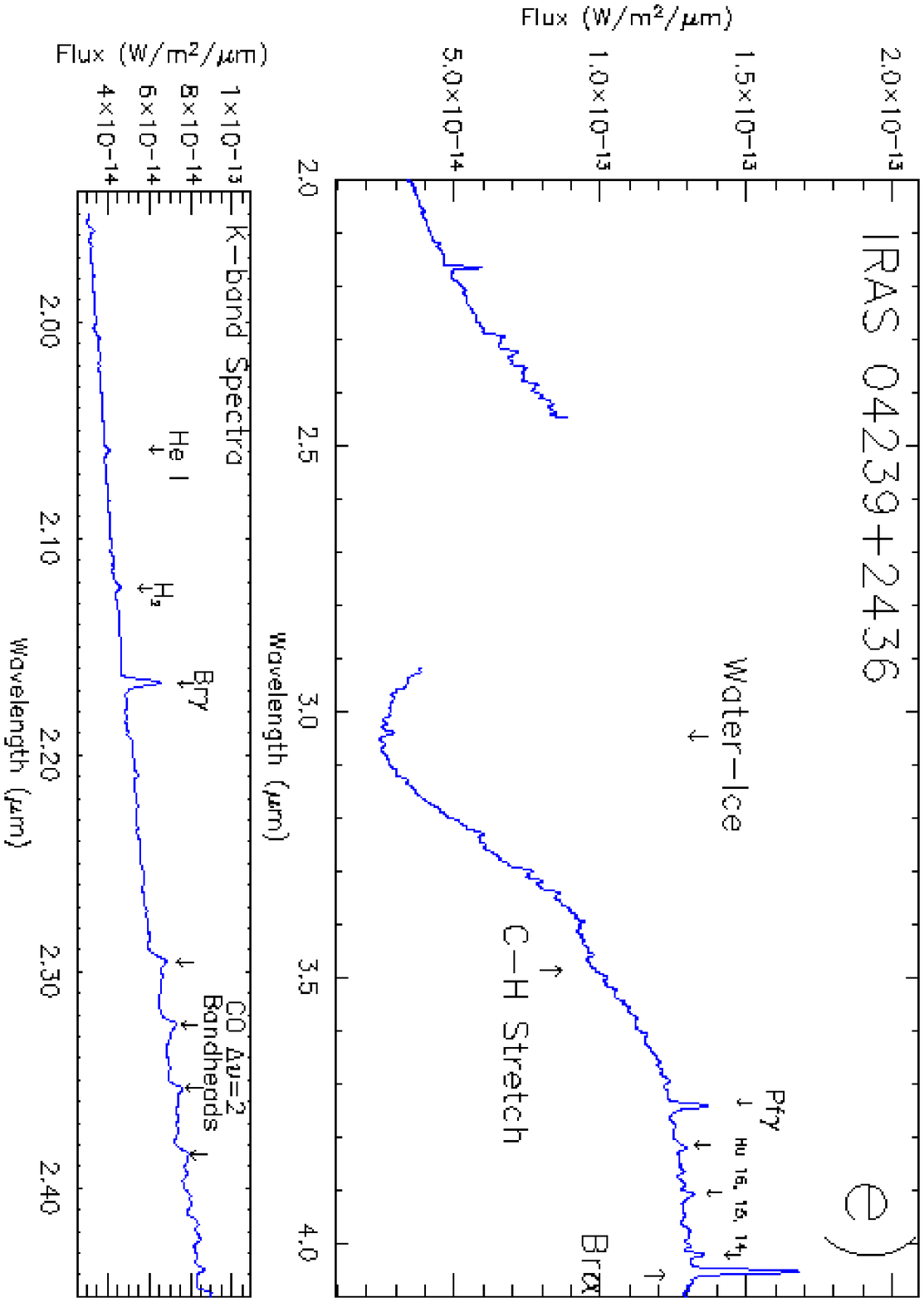}{350pt}{90pt}{40pt}{40pt}{10pt}{100pt}
\plotfiddle{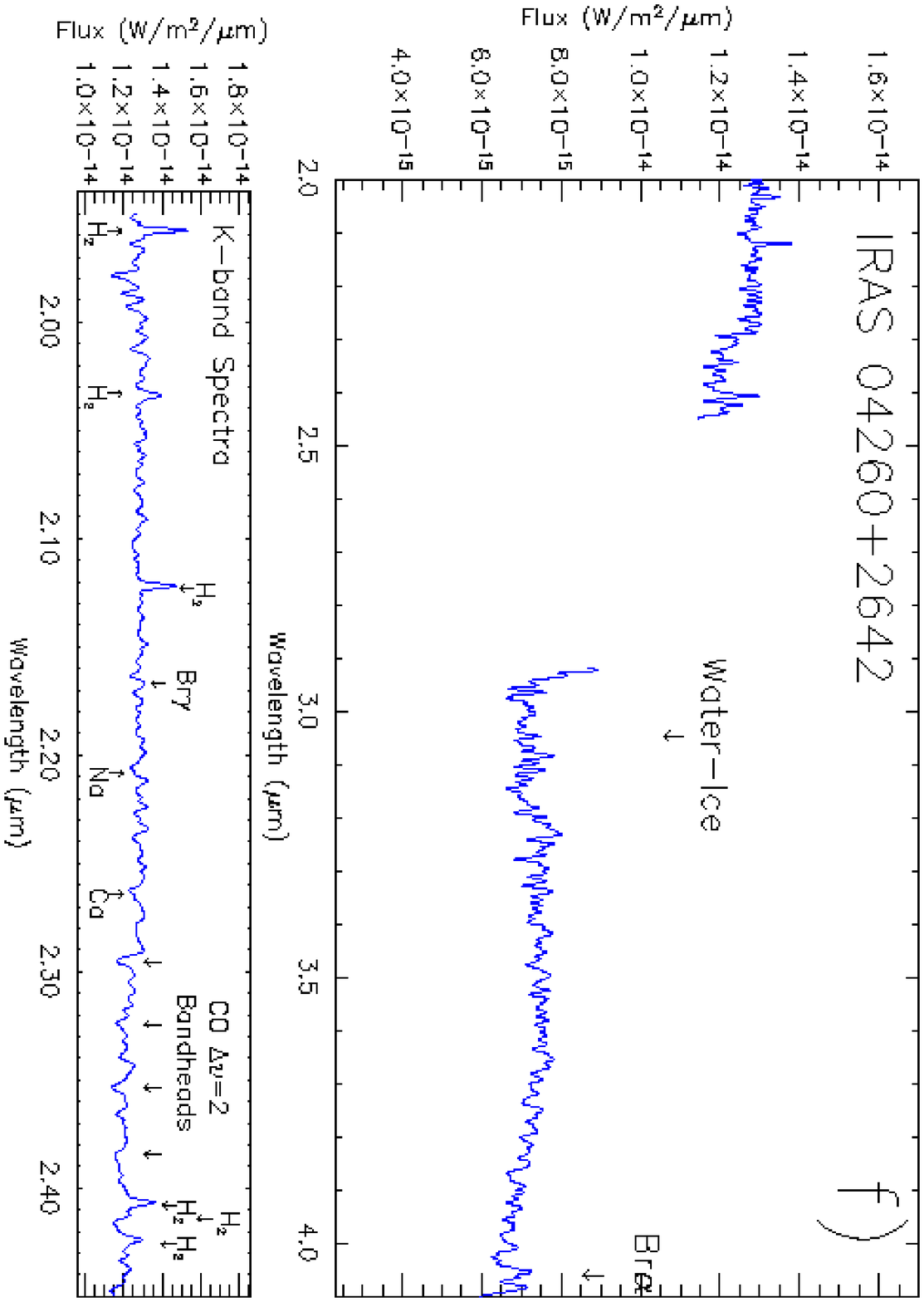}{350pt}{90pt}{40pt}{40pt}{290pt}{466pt}
\plotfiddle{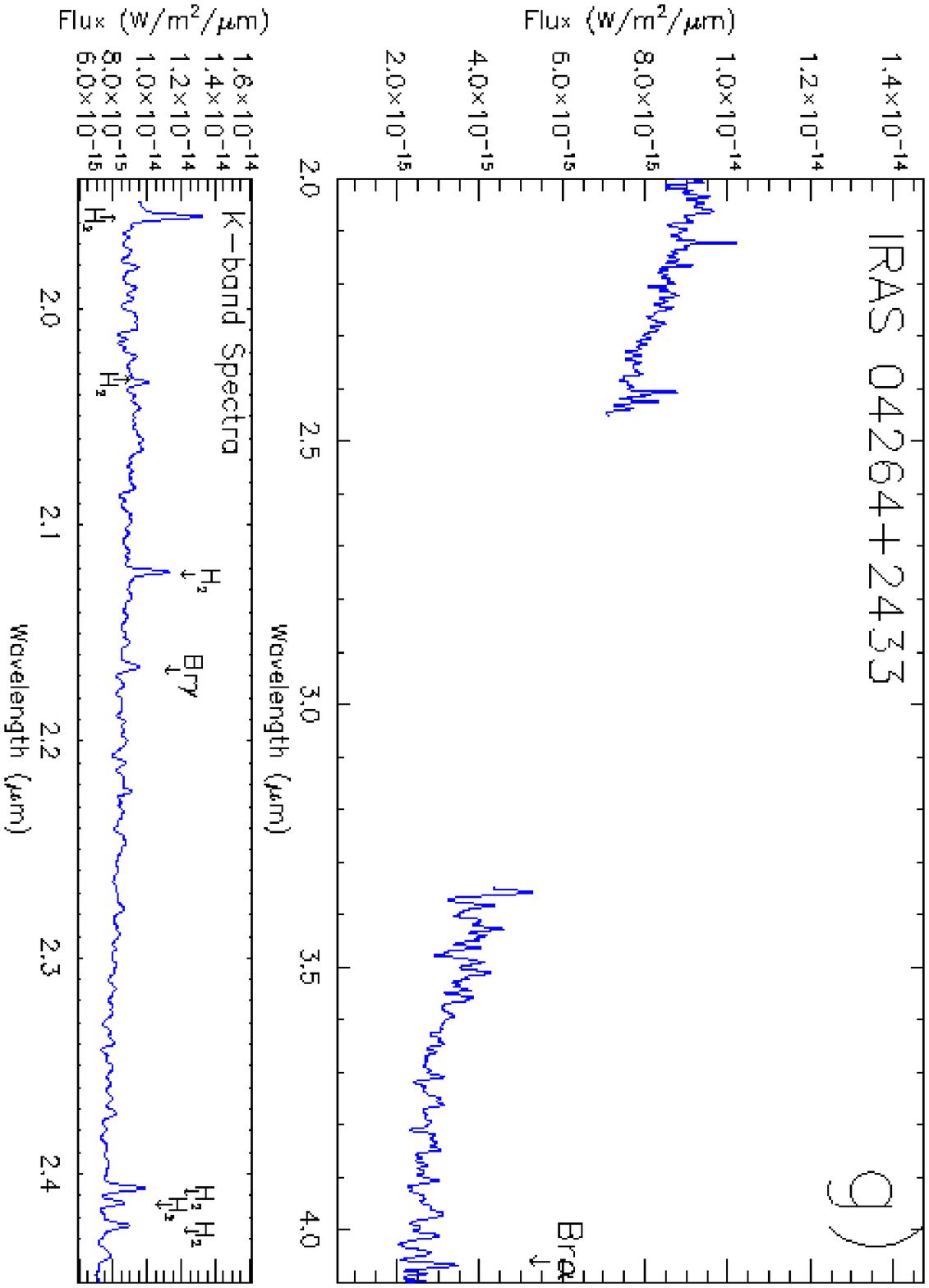}{350pt}{90pt}{40pt}{40pt}{10pt}{611pt}
\plotfiddle{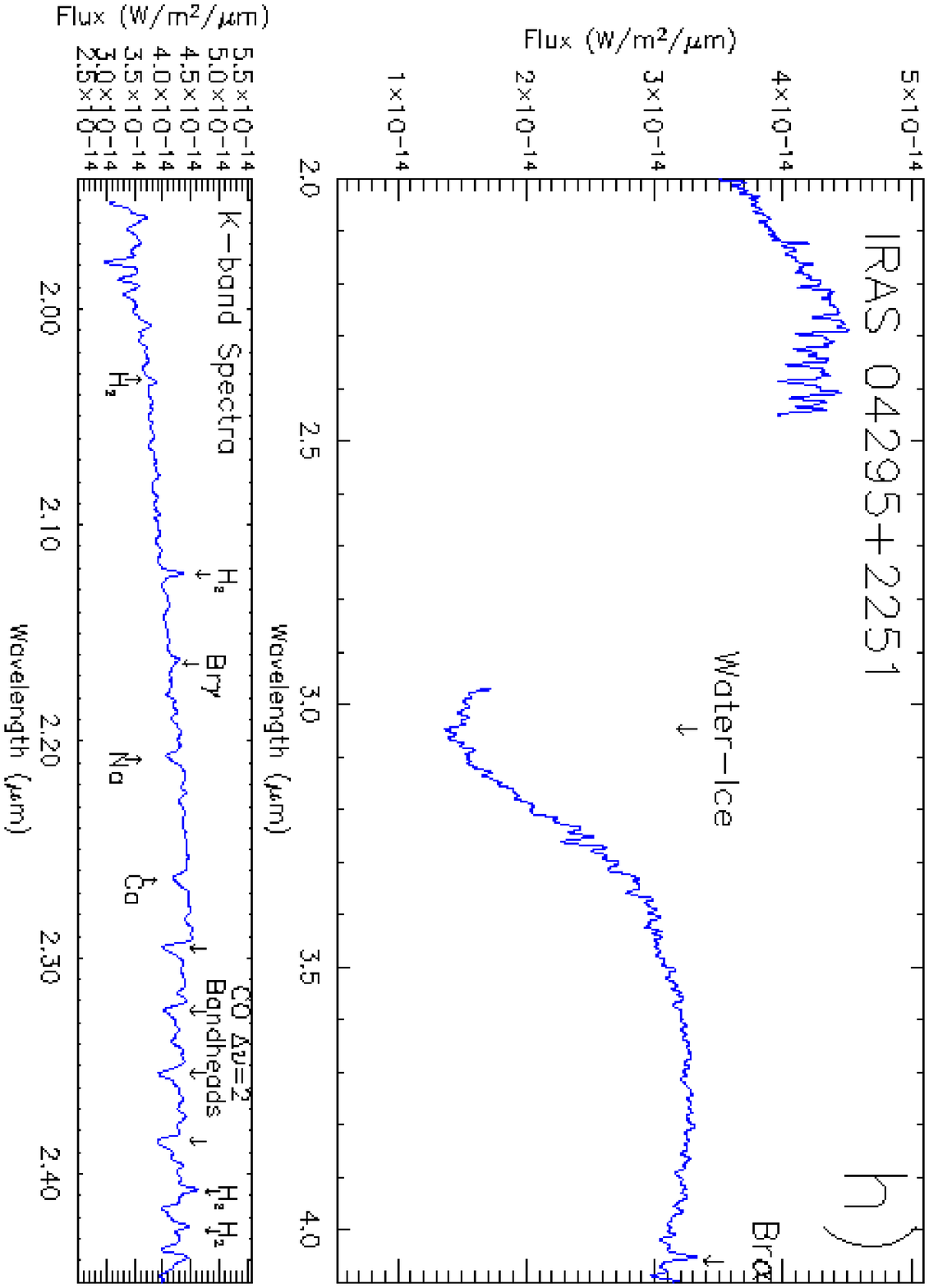}{350pt}{90pt}{40pt}{40pt}{290pt}{978pt}
\end{figure}
\newpage
\begin{figure}
\plotfiddle{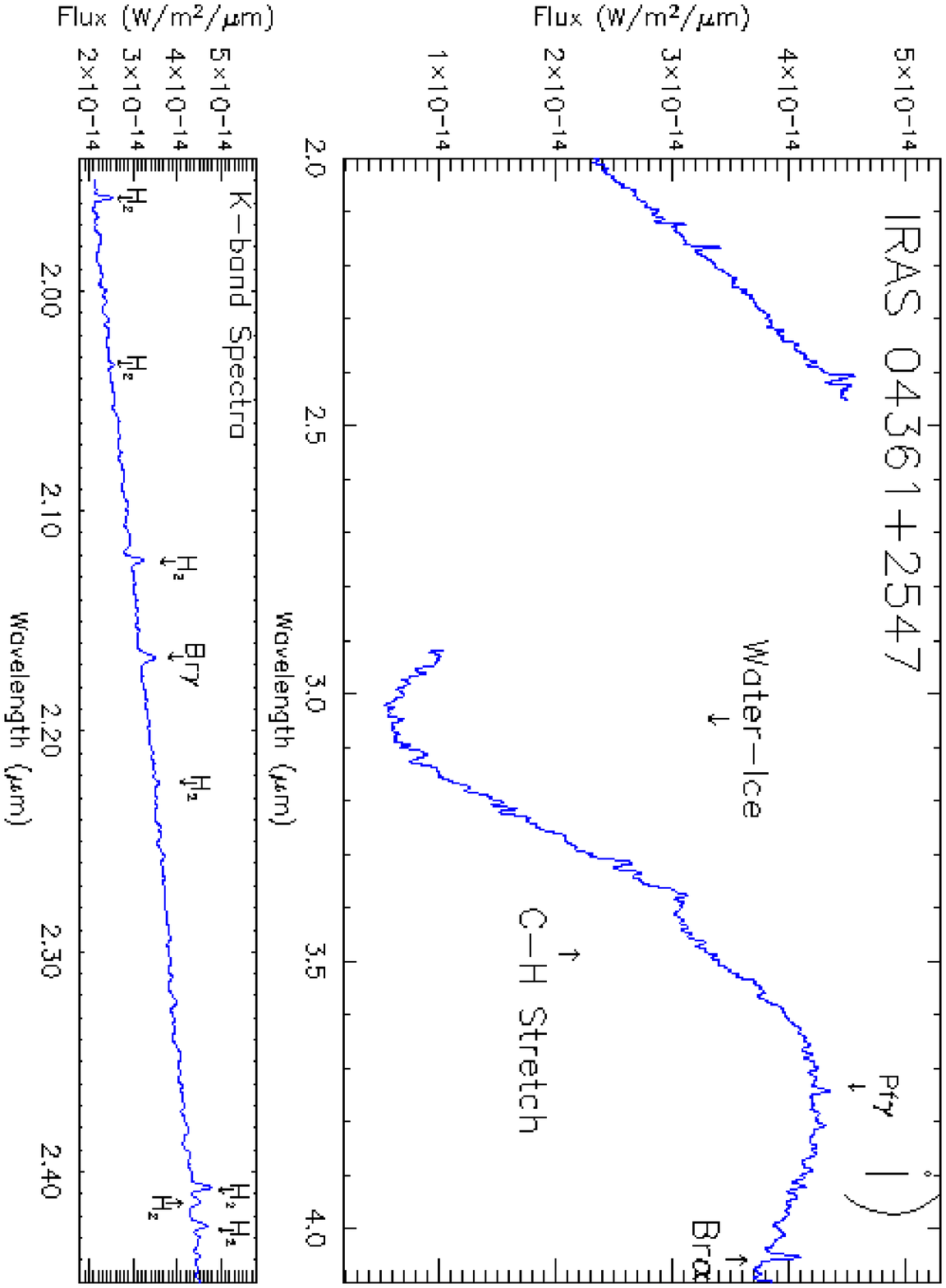}{350pt}{90pt}{40pt}{40pt}{10pt}{100pt}
\plotfiddle{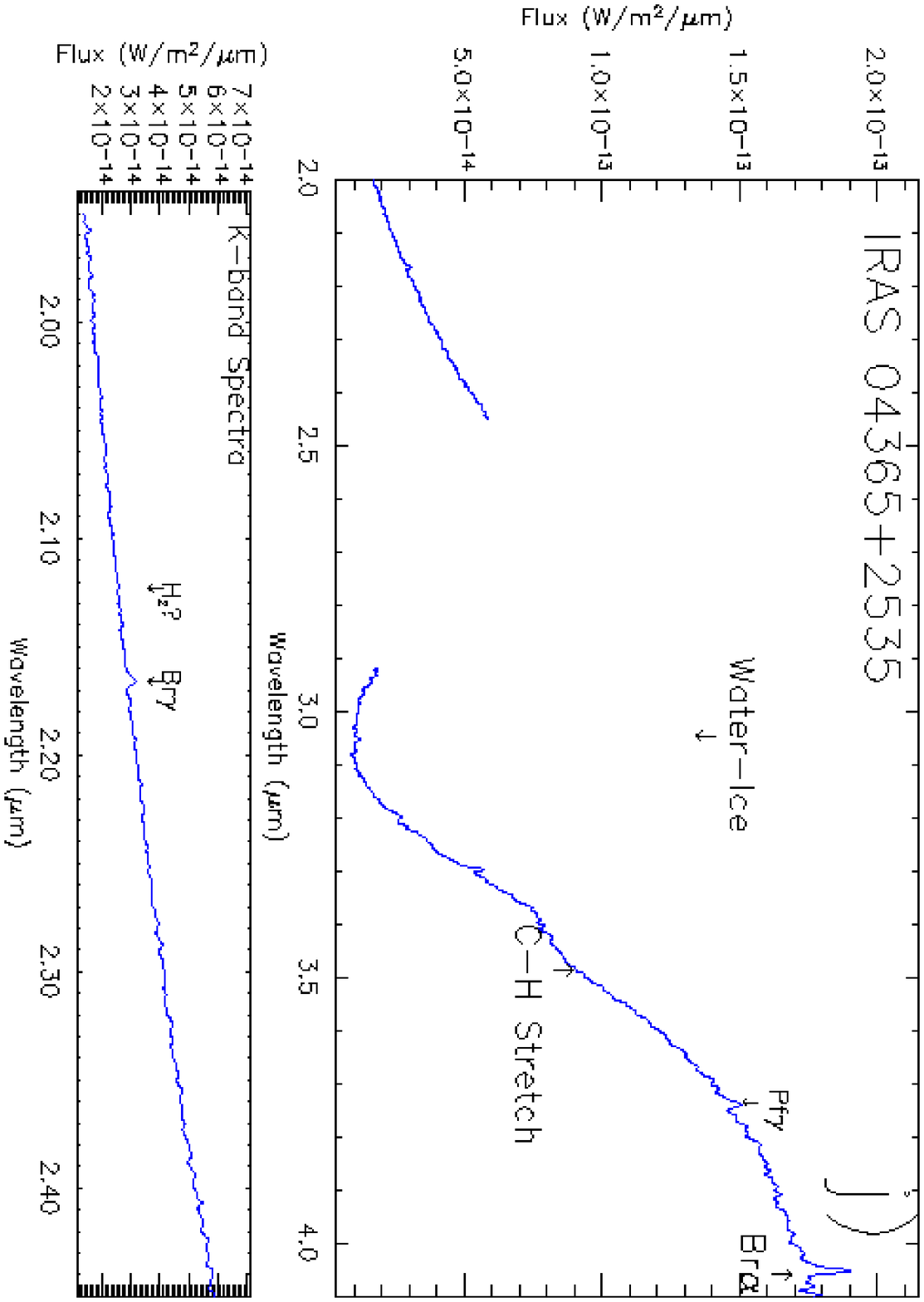}{350pt}{90pt}{40pt}{40pt}{290pt}{466pt}
\label{fig1}
\caption{}
\end{figure}

\clearpage

\begin{figure}
\plotfiddle{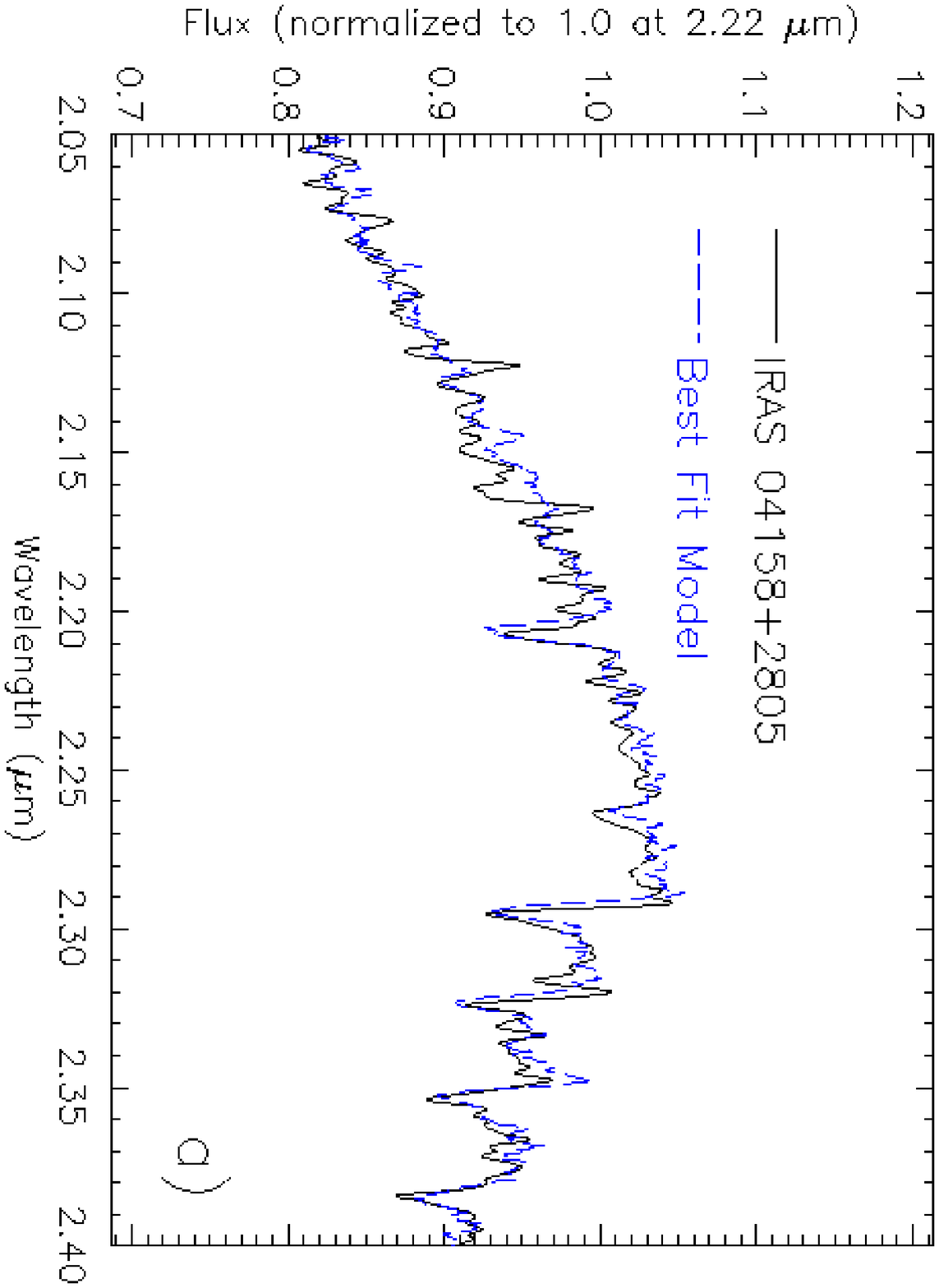}{350pt}{90pt}{40pt}{40pt}{10pt}{160pt}
\plotfiddle{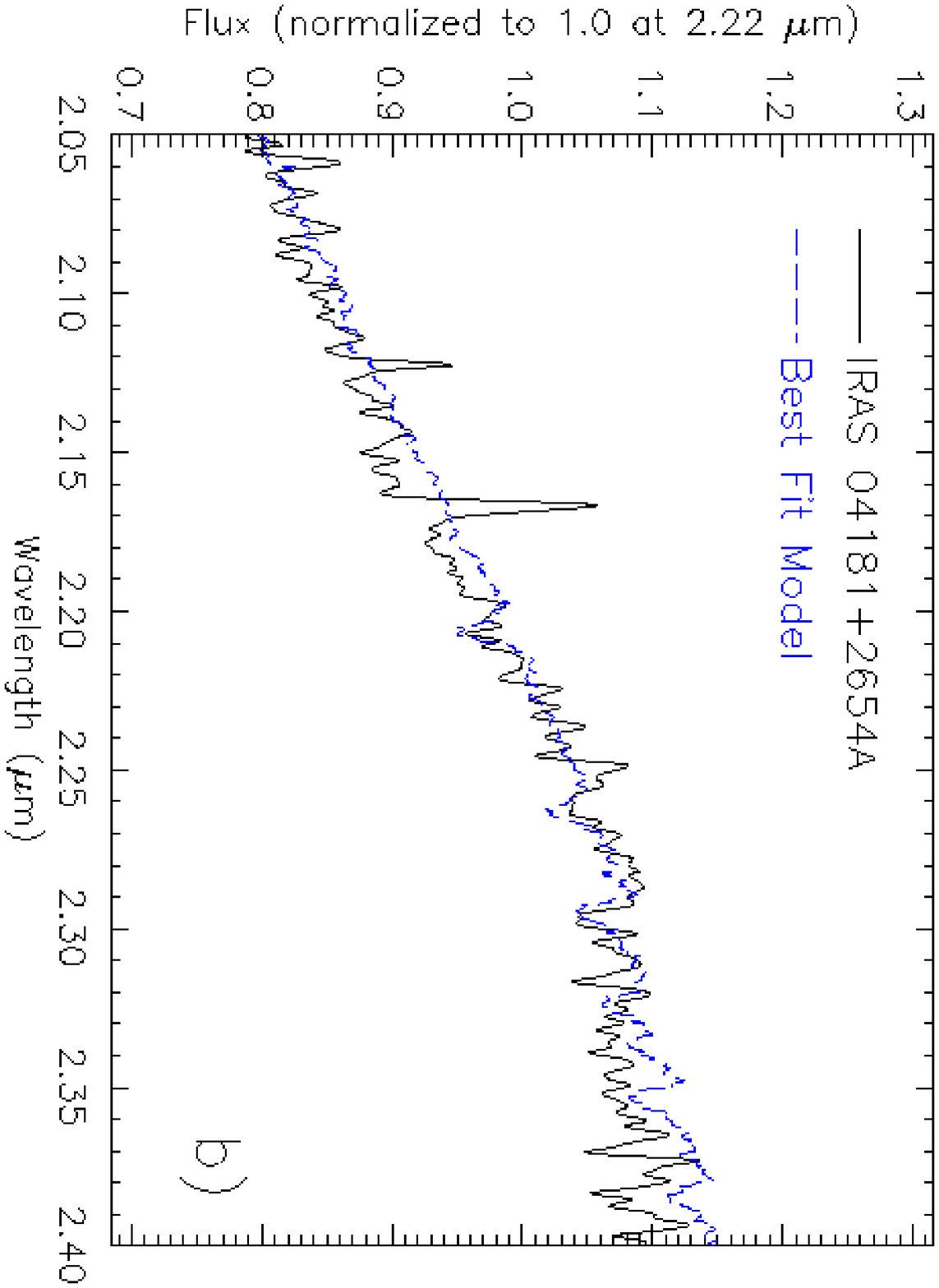}{350pt}{90pt}{40pt}{40pt}{290pt}{526pt}
\plotfiddle{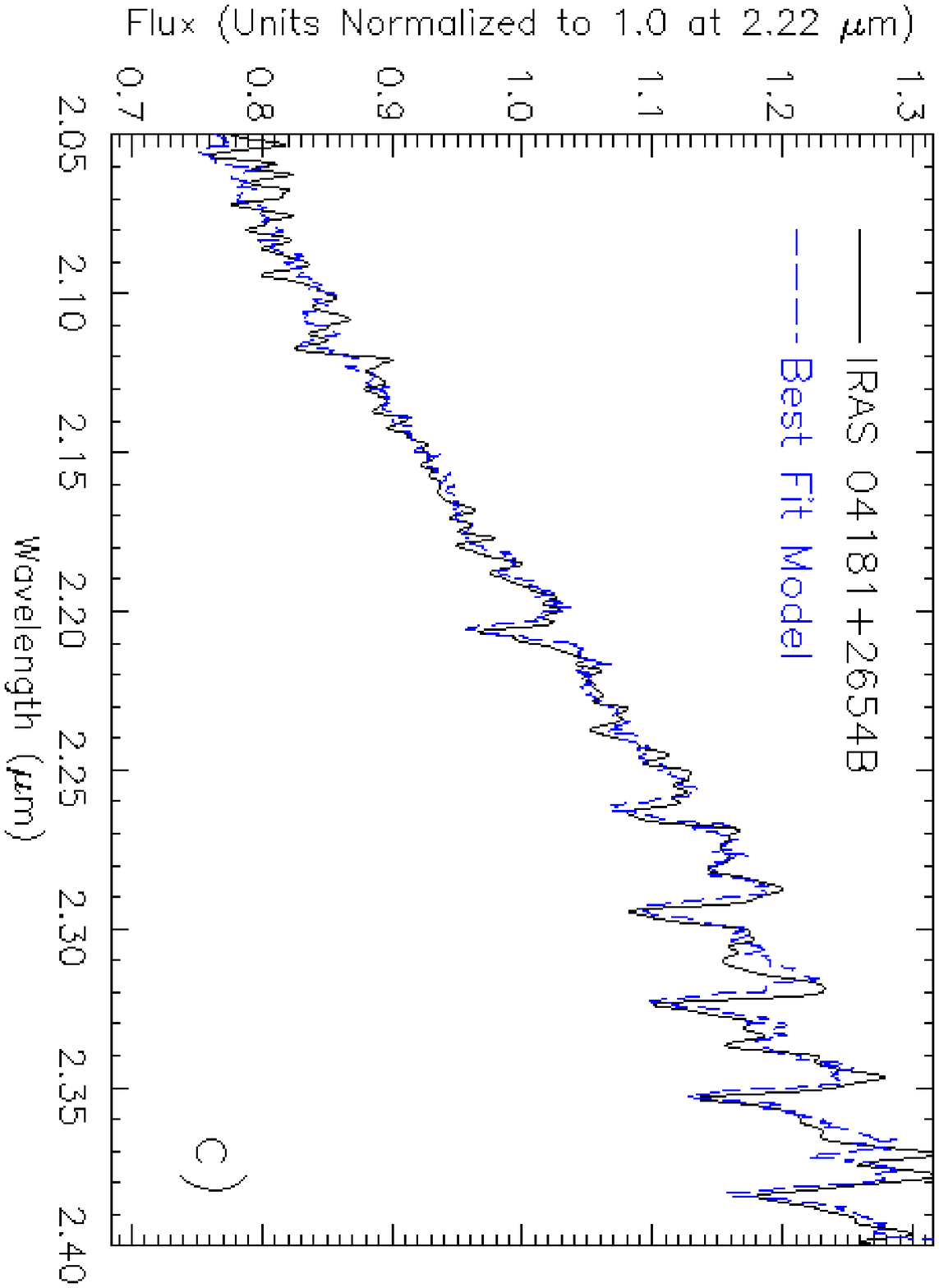}{350pt}{90pt}{40pt}{40pt}{10pt}{671pt}
\plotfiddle{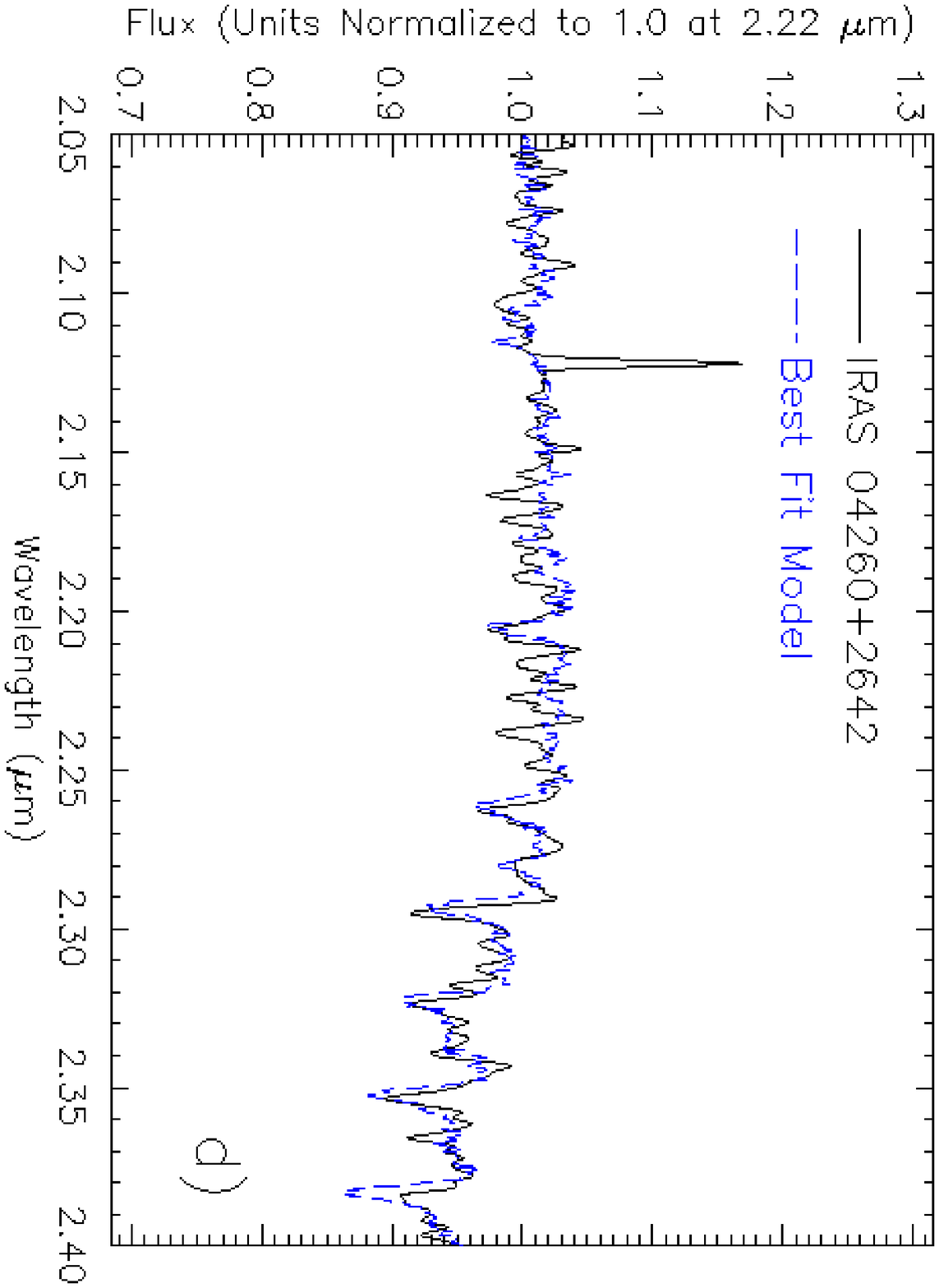}{350pt}{90pt}{40pt}{40pt}{290pt}{1038pt}
\plotfiddle{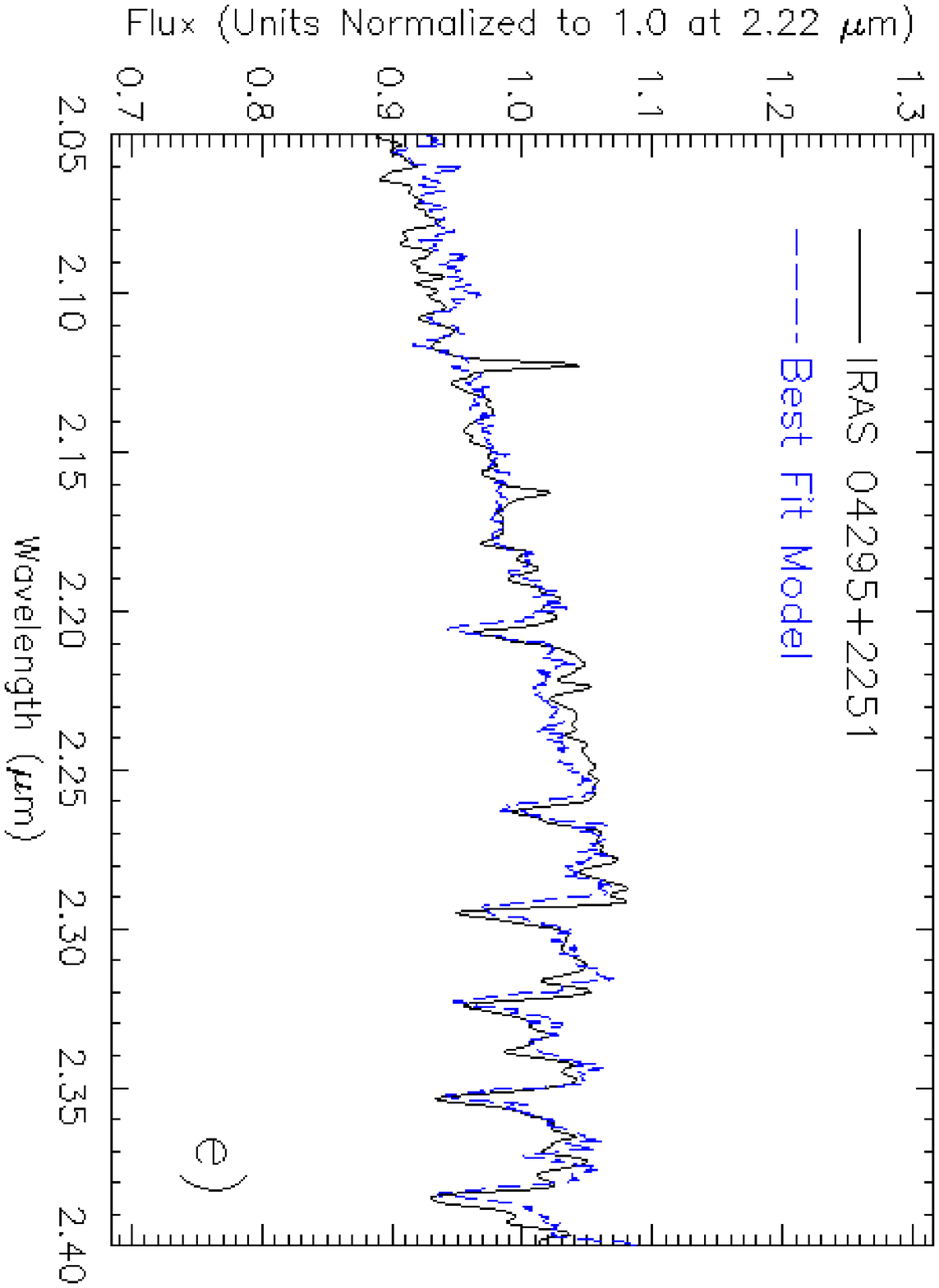}{350pt}{90pt}{40pt}{40pt}{10pt}{1183pt}
\label{fig2}
\caption{}
\end{figure}

\clearpage

\begin{figure}
\plotfiddle{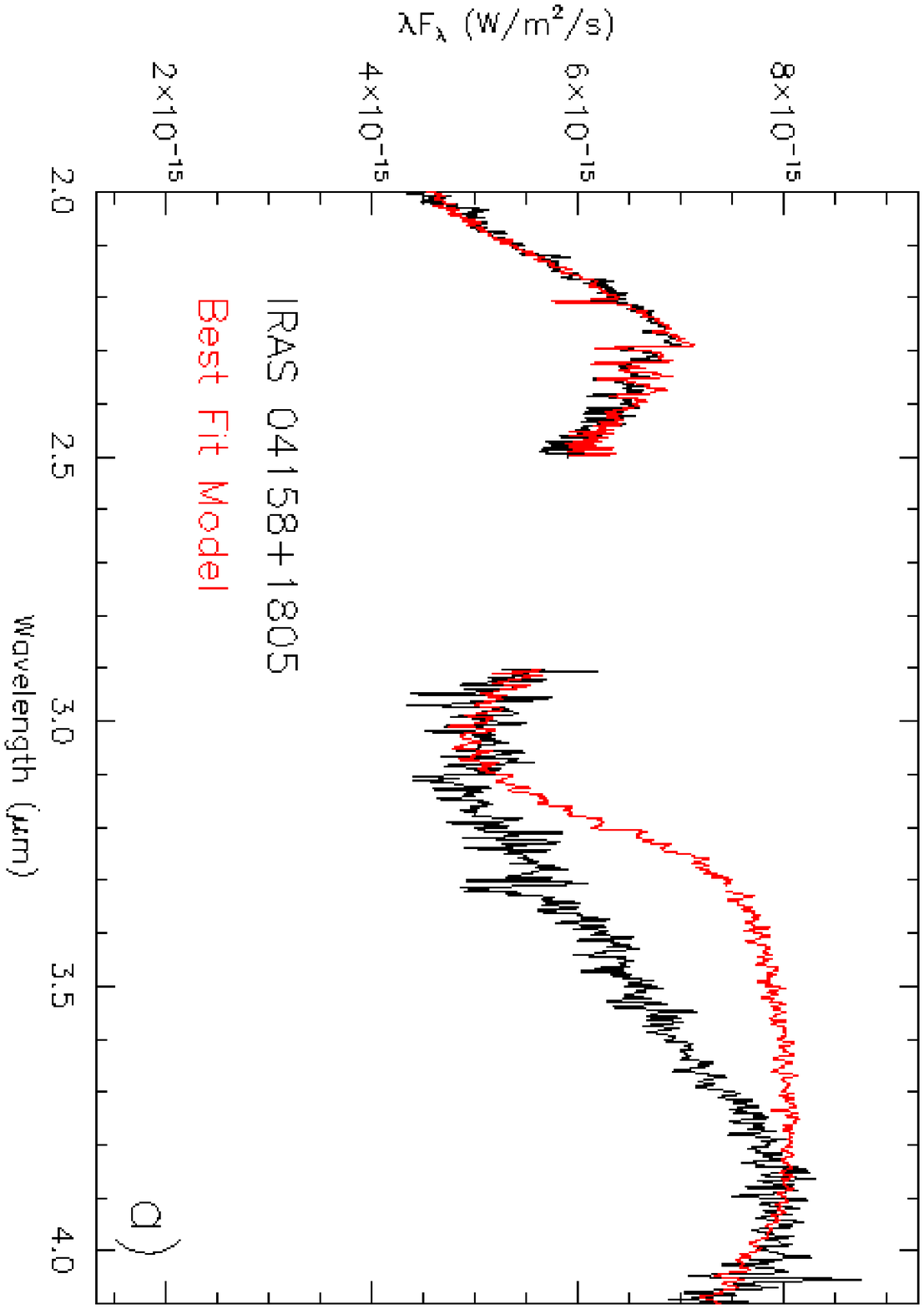}{350pt}{90pt}{40pt}{40pt}{10pt}{160pt}
\plotfiddle{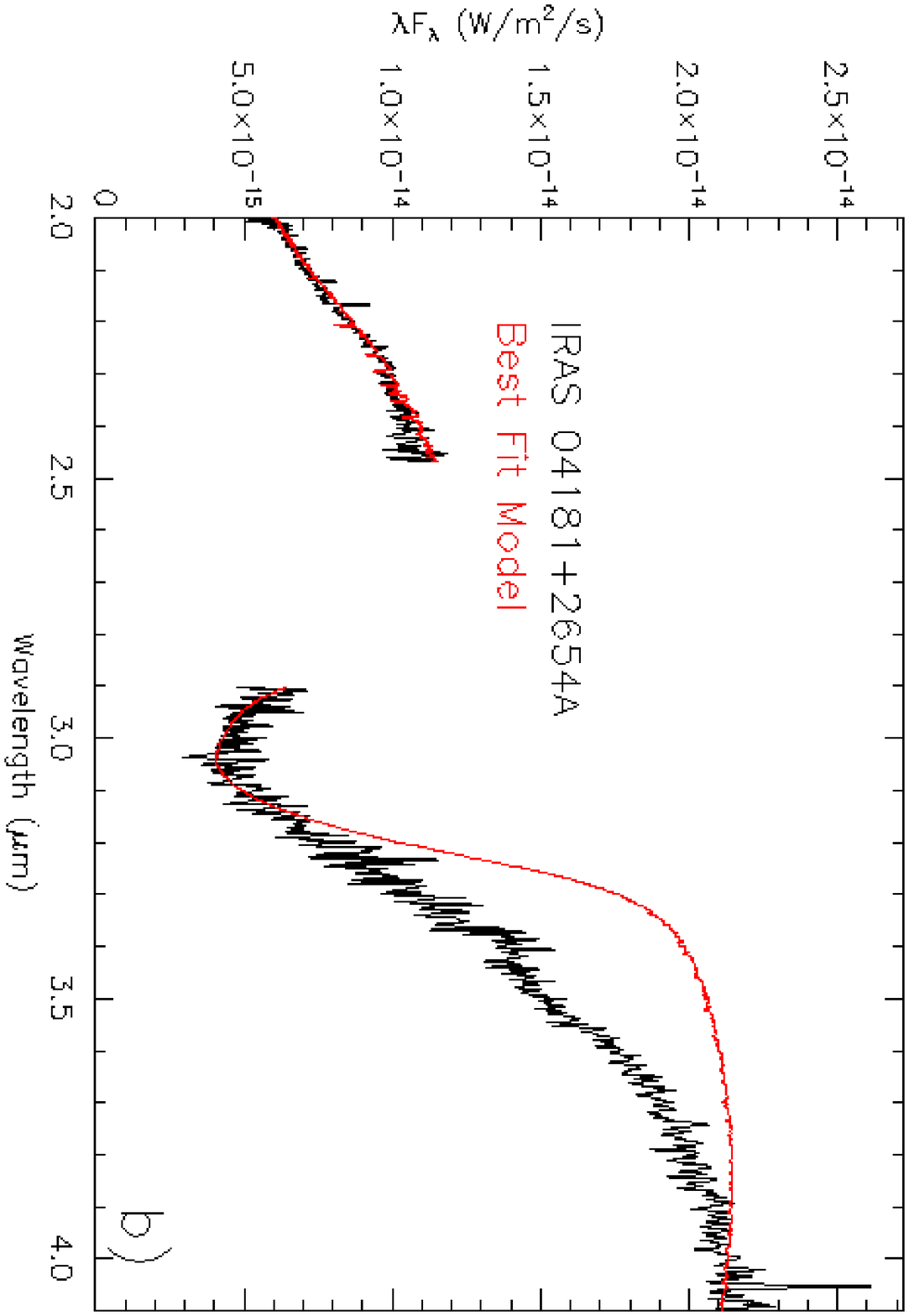}{350pt}{90pt}{40pt}{40pt}{290pt}{526pt}
\plotfiddle{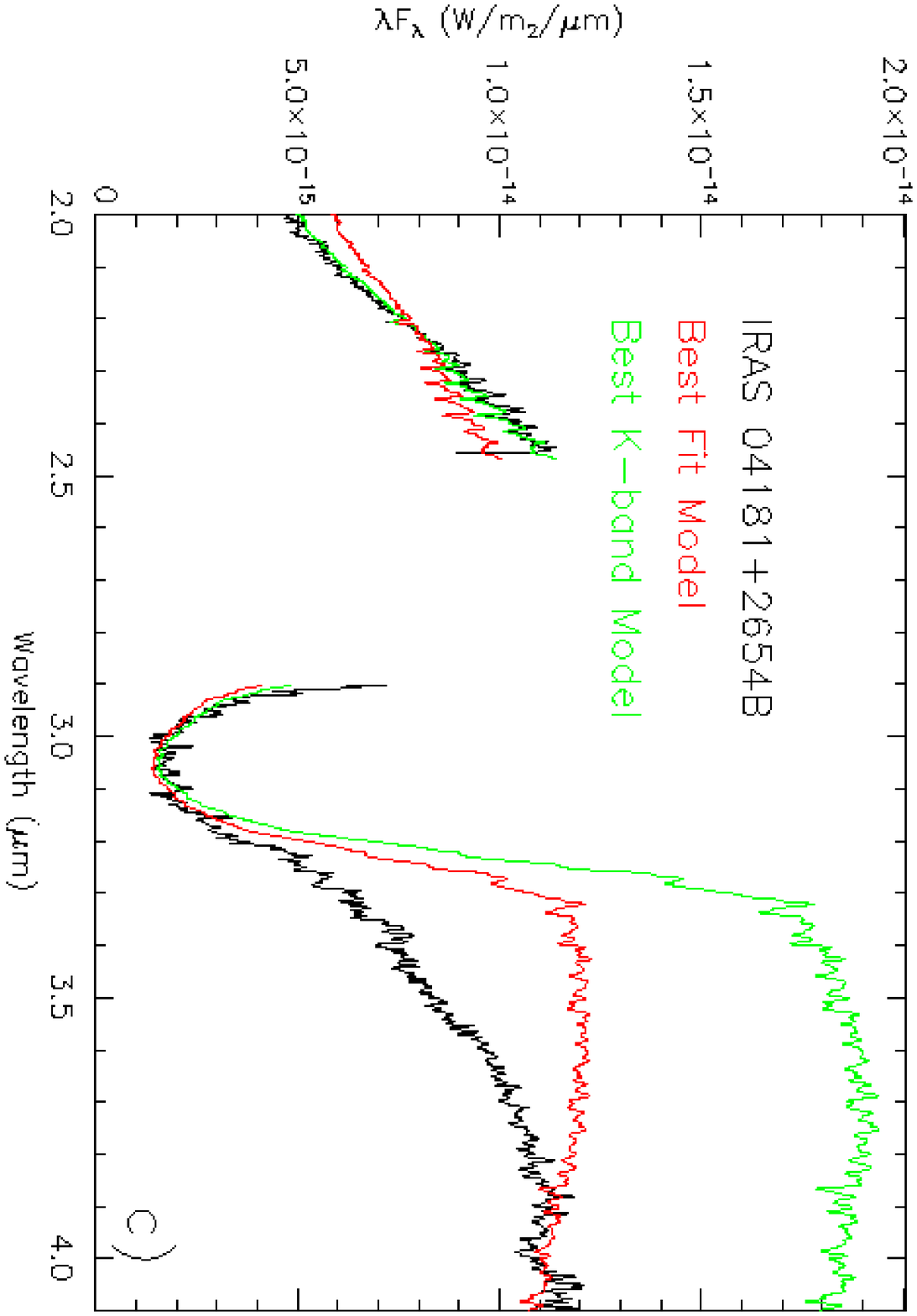}{350pt}{90pt}{40pt}{40pt}{10pt}{671pt}
\plotfiddle{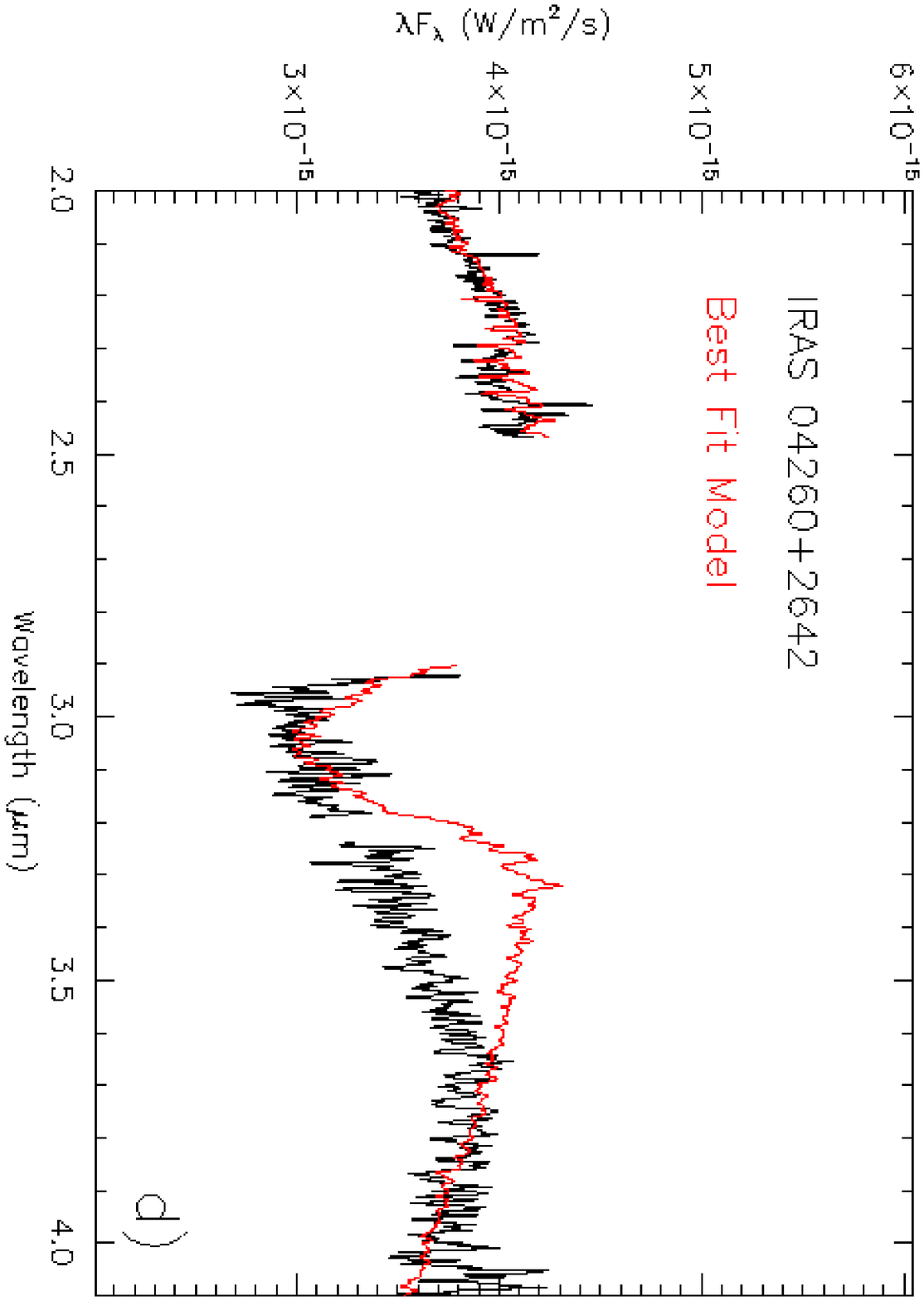}{350pt}{90pt}{40pt}{40pt}{290pt}{1038pt}
\plotfiddle{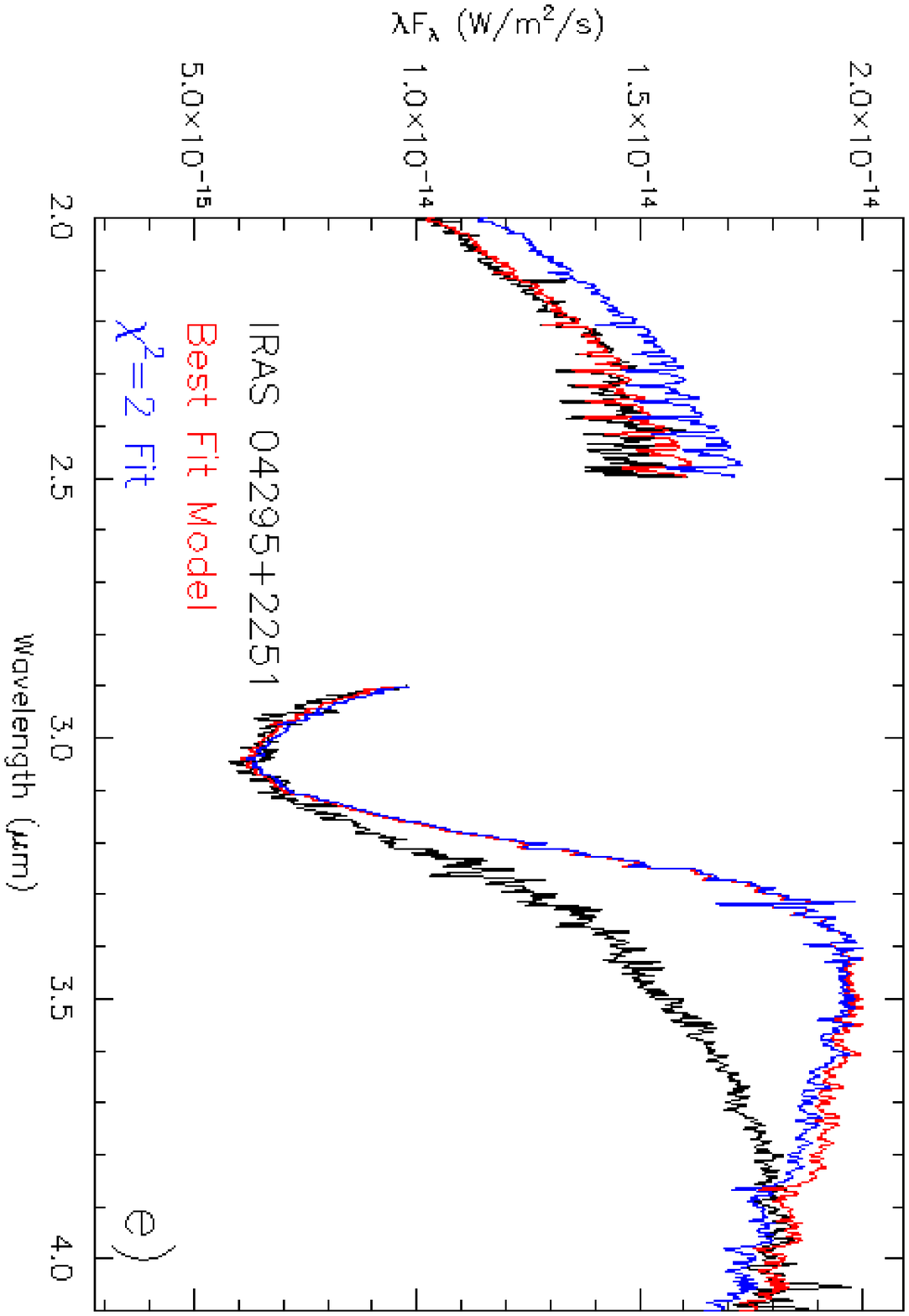}{350pt}{90pt}{40pt}{40pt}{10pt}{1183pt}
\label{fig3}
\caption{}
\end{figure}


\clearpage



\begin{figure}
\epsscale{0.7}
\includegraphics[angle=90,width=17.0cm,height=13.0cm]{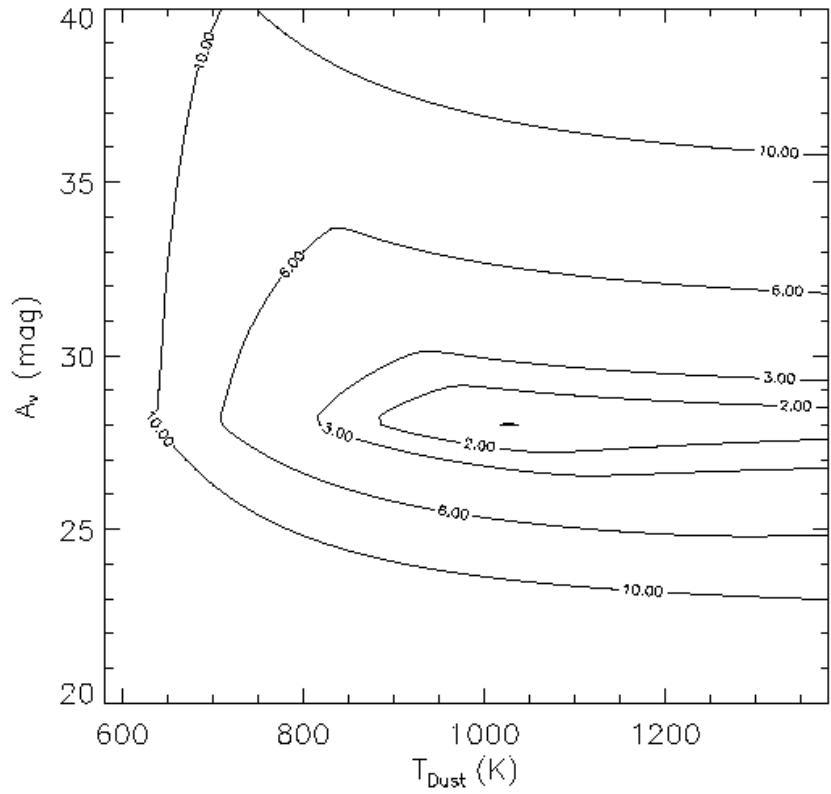}
\caption{The $\chi^2$ surface for the model fits for A$_v$ and T$_{dust}$ for IRAS 04295+2251.  Plotted are contours for $\chi^2$=1.01, 2.0, 3.0, 6.0 and 10.0.}
\label{fig4}
\end{figure}

\begin{figure}
\epsscale{0.7}
\includegraphics[angle=90,width=17.0cm,height=13.0cm]{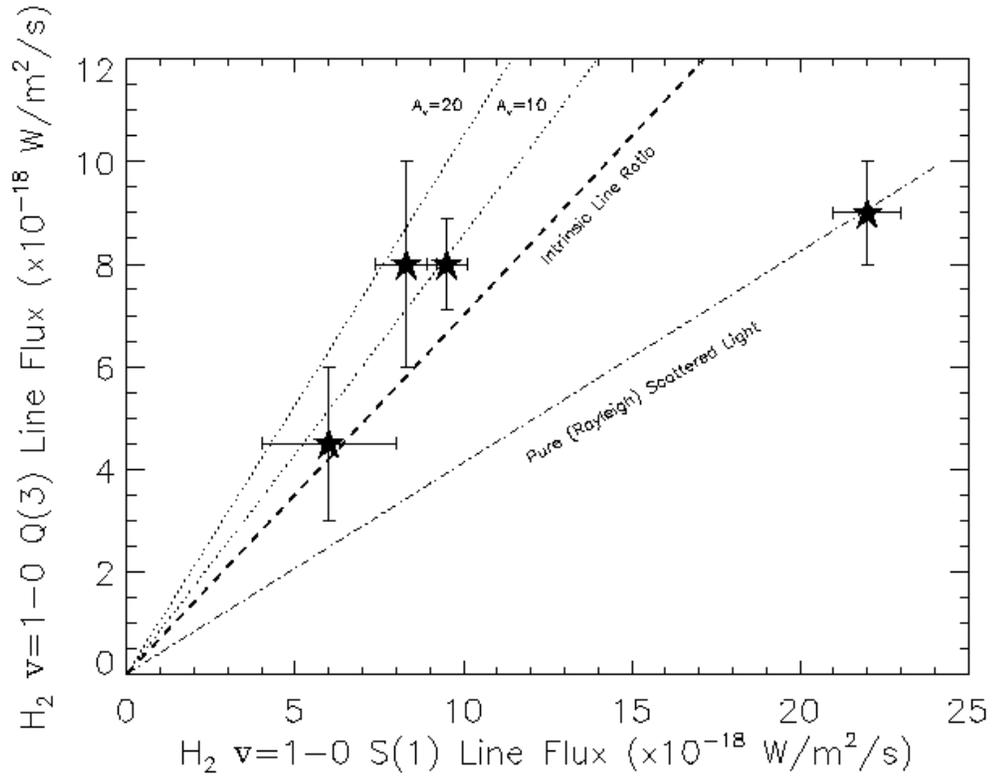}
\caption{ A comparison of the emission lines of H$_2$ {\it v}=1-0 Q(3) versus  H$_2$ {\it v}=1-0 S(1).  The intrinsic line ratio is 0.7; deviations from this value are caused by either extinction or scattering.  Overplotted in the figure are the resulting line ratios from A$_{v}$=10mag, 20 mag of extinction, and a scenario where the molecular hydrogen flux is unobscured but seen purely in scattered light with a Rayleigh wavelength dependence.}
\label{fig5}
\end{figure}

\clearpage

\begin{figure}
\epsscale{0.7}
\includegraphics[angle=90,width=17.0cm,height=13.0cm]{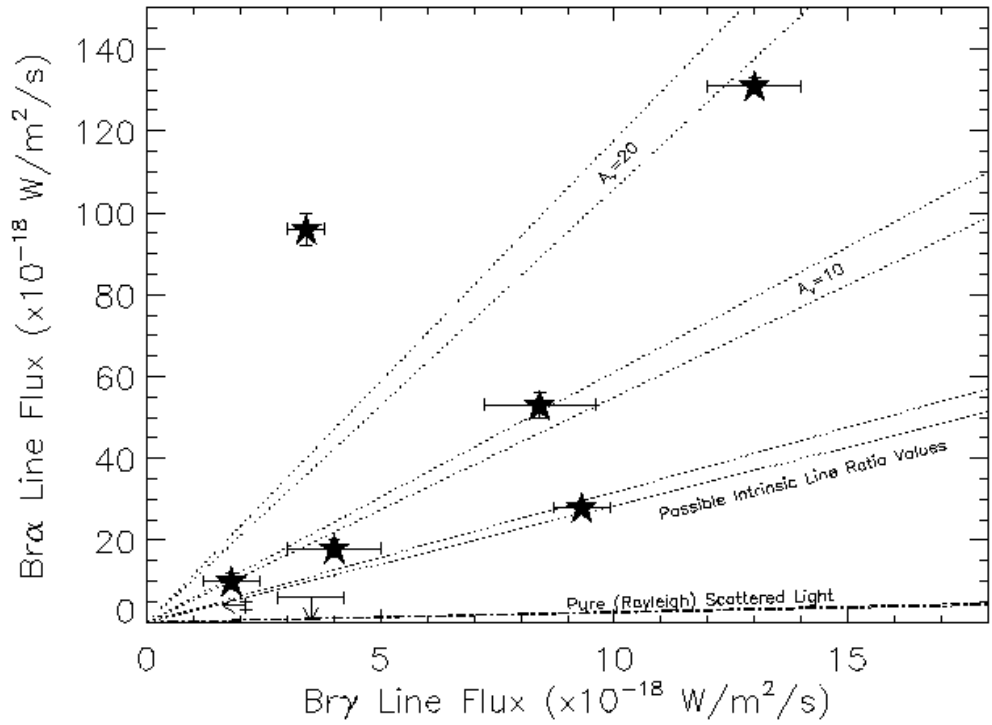}
\caption{  A comparison of the emission features of Br$\alpha$ plotted versus Br$\gamma$.  Intrinsic line ratios in a Baker \& Menzel (1938) Case B treatment are in the range of 2.85 to 3.17 (Brockelhurst 1971); deviations from this range are presumably caused by either extinction or scattering.  Overplotted in the figure are the ranges that would result from A$_{v}$=10mag extinction, 20 mag of extinction, and a scenario where the Brackett fluxes are seen purely in scattered light, as in Figure 5. IRAS 04239+2436 is outside to the upper right of the plotted limits.}
\label{fig6}
\end{figure}

\clearpage

\begin{figure}
\epsscale{0.6}
\includegraphics[angle=90,width=19.0cm,height=14.5cm]{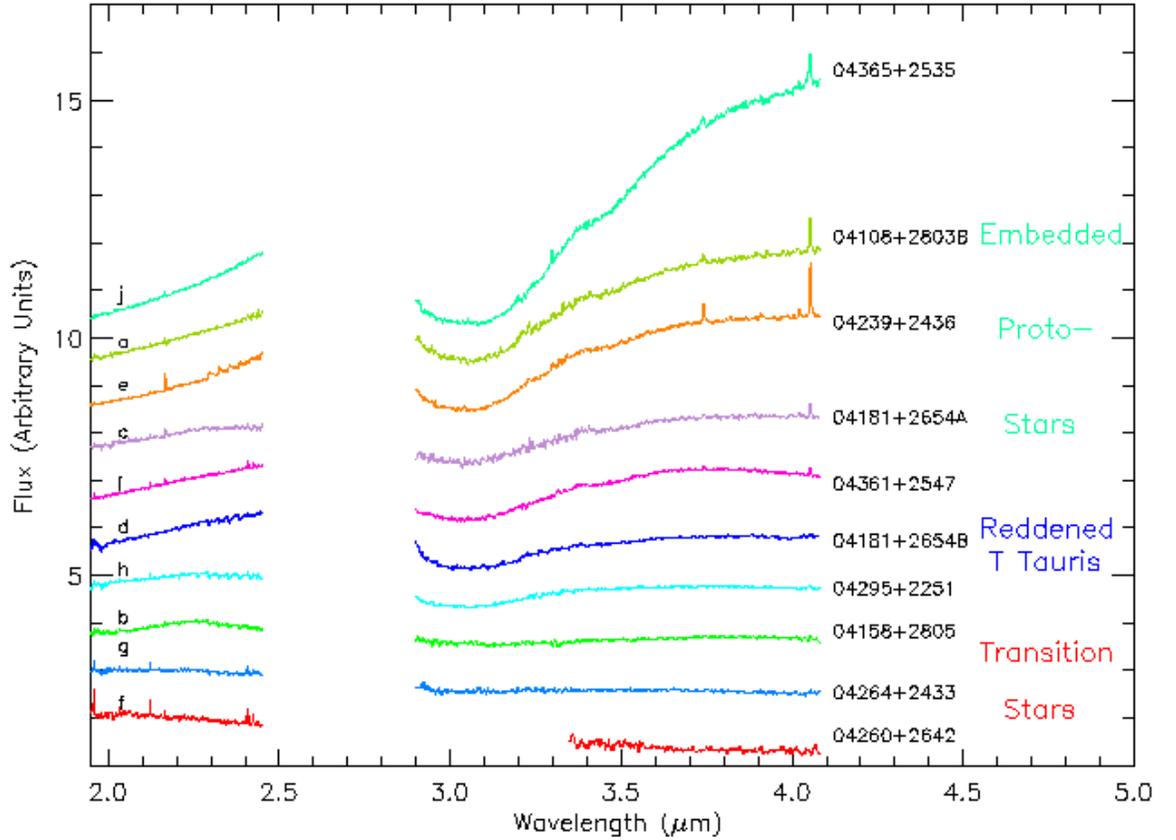}
\caption{The 2-4$\mu$m spectra for all 10 targets from this study, plotted from bottom to top in order of increasing K-L$'$ color.  The lower stars in the plot are best described as transition Class I/Class II stars, and the upper are true Class I protostars.    Included to the left of the figure is a note on the order of the spectra from Figure 1.  IRAS 04181+2654B and IRAS 04295+2251 could be more evolved Class II T Tauri stars that are heavily obscured (see \S6).}
\label{fig7}
\end{figure}
\clearpage




\end{document}